\documentclass[usenatbib]{mnras_arxiv} 
\usepackage{txfonts} 
\usepackage[T1]{fontenc}
\usepackage{ae,aecompl}
\usepackage{natbib}
\usepackage{fixltx2e}
\usepackage{graphicx}
\usepackage{epstopdf} 
\usepackage{epsfig}
\usepackage{float}
\usepackage{amssymb}
\usepackage{color}
\usepackage{caption}
\usepackage{subcaption} 
\usepackage{aecompl}
\usepackage[figuresleft]{rotating}
\usepackage{afterpage}
\usepackage{longtable}
\usepackage{appendix}
\newcommand{\affil}[1]{$^{\rm #1}$}
\newcounter{inst} 
\newcommand{\inst}[1]{\noindent%
   \refstepcounter{inst}\affil{\arabic{inst}\label{#1}}     
   }
\newcounter{ft}
\newcounter{todo}
\renewcommand\thetodo{\Alph{todo}}
\def\todo#1{\addtocounter{todo}{1}[[\thetodo: #1]]\strut\vadjust{%
\kern-\dp\strutbox{\vtop to \dp\strutbox{\baselineskip\dp\strutbox\vss\rlap{%
\hskip\hsize\ \rm{$\leftarrow$\thetodo}}\null}}}}
\def\LaTeX{L\kern-.36em\raise.3ex\hbox{a}\kern-.15em 
    T\kern-.1667em\lower.7ex\hbox{E}\kern-.125emX}

\newcommand{\arcdeg}{\ensuremath{^{\circ}}}
\newcommand{\survarea}{24,831 }
\newcommand{\npos}{245,457 }
\newcommand{\nfit}{238,364 }
\newcommand{\nsrc}{307,455 }
\newcommand{\nresolved}{97,103 }

\newcommand{\nclustered}{90,237 }
\newcommand{\ncol}{311 }
\newcommand{\pctreliable}{99.97}

\def\fig{Figure}

\def\Fig{Figure}

\def\sect{Section}

\def\Sect{Section}
\def\Sects{Sections}

\def\Tab{Table}

\title[GLEAM Survey I: Extragalactic catalogue]{GaLactic and Extragalactic All-sky Murchison Widefield Array (GLEAM) survey I: A low-frequency extragalactic catalogue}
\author[Hurley-Walker~et~al.]{N.~Hurley-Walker\affil{\ref{ICRAR}},
J.~R.~Callingham\affil{\ref{USyd},\ref{CASS},\ref{CAASTRO}},
P.~J.~Hancock\affil{\ref{ICRAR},\ref{CAASTRO}}, 
T.~M.~O.~Franzen\affil{\ref{ICRAR}},\newauthor
L.~Hindson\affil{\ref{VUW}},
A.~D.~Kapi\'nska\affil{\ref{CAASTRO},\ref{UWA}},
J.~Morgan\affil{\ref{ICRAR}},
A.~R.~Offringa\affil{\ref{CAASTRO},\ref{ASTRON}},
R.~B.~Wayth\affil{\ref{ICRAR},\ref{CAASTRO}},  \newauthor
C.~Wu\affil{\ref{UWA}},
Q.~Zheng\affil{\ref{VUW}},
T.~Murphy\affil{\ref{USyd},\ref{CAASTRO}},
M.~E.~Bell\affil{\ref{CASS},\ref{CAASTRO}},  
K.~S.~Dwarakanath\affil{\ref{RRI}}, 
B.~For\affil{\ref{UWA}}, \newauthor
B.~M.~Gaensler\affil{\ref{USyd},\ref{CAASTRO},\ref{Toronto}},
M.~Johnston-Hollitt\affil{\ref{VUW}}, 
E.~Lenc\affil{\ref{USyd},\ref{CAASTRO}},
P.~Procopio\affil{\ref{CAASTRO},\ref{UMelb}},\newauthor
L.~Staveley-Smith\affil{\ref{UWA}}, 
R.~Ekers\affil{\ref{ICRAR},\ref{USyd},\ref{CAASTRO}}
J.~D.~Bowman\affil{\ref{ASU}}, 
F.~Briggs\affil{\ref{CAASTRO},\ref{ANU}},  
R.~J.~Cappallo\affil{\ref{Haystack}},\newauthor
A.~A.~Deshpande\affil{\ref{RRI}}, 
L.~Greenhill\affil{\ref{CfA}},
B.~J.~Hazelton\affil{\ref{UDub}}, 
D.~L.~Kaplan\affil{\ref{UWM}},
C.~J.~Lonsdale\affil{\ref{Haystack}},\newauthor 
S.~R.~McWhirter\affil{\ref{Haystack}}, 
D.~A~Mitchell\affil{\ref{CASS},\ref{CAASTRO}}, 
M.~F.~Morales\affil{\ref{UDub}},
E.~Morgan\affil{\ref{MIT}},
D.~Oberoi\affil{\ref{NCRA}},\newauthor
S.~M.~Ord\affil{\ref{ICRAR},\ref{CAASTRO}}, 
T.~Prabu\affil{\ref{RRI}}, 
N.~Udaya~Shankar\affil{\ref{RRI}},
K.~S.~Srivani\affil{\ref{RRI}},
R.~Subrahmanyan\affil{\ref{CAASTRO},\ref{RRI}}, \newauthor
S.~J.~Tingay\affil{\ref{ICRAR},\ref{CAASTRO}},
R.~L.~Webster\affil{\ref{CAASTRO},\ref{UMelb}}, 
A.~Williams\affil{\ref{ICRAR}},
C.~L.~Williams\affil{\ref{MIT}}\\
{\small \affil{}\,Email: nhw@icrar.org}\\
{\small \inst{ICRAR}\,International Centre for Radio Astronomy Research, Curtin University, Bentley, WA 6102, Australia}\\
{\small \inst{USyd}\,Sydney Institute for Astronomy, School of Physics, The University of Sydney, NSW 2006, Australia}\\
{\small \inst{CASS}\,CSIRO Astronomy and Space Science, Marsfield, NSW 2122, Australia}\\
{\small \inst{CAASTRO}\,ARC Centre of Excellence for All-sky Astrophysics (CAASTRO)}\\
{\small \inst{VUW}\,School of Chemical \& Physical Sciences, Victoria University of Wellington, Wellington 6140, New Zealand}\\
{\small \inst{UWA}\,International Centre for Radio Astronomy Research, University of Western Australia, Crawley 6009, Australia}\\
{\small \inst{ASTRON}\,Netherlands Institute for Radio Astronomy (ASTRON), PO Box 2, 7990 AA Dwingeloo, The Netherlands}\\
{\small \inst{RRI}\,Raman Research Institute, Bangalore 560080, India}\\
{\small \inst{Toronto}\,Dunlap Institute for Astronomy and Astrophysics, 50 St. George St, University of Toronto, ON M5S 3H4, Canada}\\
{\small \inst{UMelb}\,School of Physics, The University of Melbourne, Parkville, VIC 3010, Australia}\\
{\small \inst{ASU}\,School of Earth and Space Exploration, Arizona State University, Tempe, AZ 85287, USA}\\
{\small \inst{ANU}\,Research School of Astronomy and Astrophysics, Australian National University, Canberra, ACT 2611, Australia}\\
{\small \inst{Haystack}\,MIT Haystack Observatory, Westford, MA 01886, USA}\\
{\small \inst{CfA}\,Harvard-Smithsonian Center for Astrophysics, 60 Garden Street, Cambridge, MA, 02138, USA}\\  
{\small \inst{UDub}\,Department of Physics, University of Washington, Seattle, WA 98195, USA}\\
{\small \inst{UWM}\,Department of Physics, University of Wisconsin--Milwaukee, Milwaukee, WI 53201, USA}\\
{\small \inst{MIT}\,Kavli Institute for Astrophysics and Space Research, Massachusetts Institute of Technology, Cambridge, MA 02139, USA}\\
{\small \inst{NCRA}\,National Centre for Radio Astrophysics, Tata Institute for Fundamental Research, Pune 411007, India}\\
}
\begin{document}

\date{Accepted 0000 December 00. Received 0000 December 00; in original form 0000 October 00}

\pagerange{\pageref{firstpage}--\pageref{lastpage}} \pubyear{2016}

\maketitle

\label{firstpage}

\begin{abstract}

Using the Murchison Widefield Array (MWA), the low-frequency Square Kilometre Array (SKA1 LOW) precursor located in Western Australia, we have completed the GaLactic and Extragalactic All-sky MWA (GLEAM) survey, and present the resulting extragalactic catalogue, utilising the first year of observations. The catalogue covers $\survarea$ square degrees, over declinations south of $+30\arcdeg$ and Galactic latitudes outside $10\arcdeg$ of the Galactic plane, excluding some areas such as the Magellanic Clouds. It contains \nsrc radio sources with 20~separate flux density measurements across 72--231\,MHz, selected from a time- and frequency- integrated image centred at 200\,MHz, with a resolution of $\approx 2\arcmin$. Over the catalogued region, we estimate that the catalogue is 90\,\% complete at 170\,mJy, and 50\,\% complete at 55\,mJy, and large areas are complete at even lower flux density levels.
Its reliability is \pctreliable\,\% above the detection threshold of $5\sigma$, which itself is typically 50\,mJy. These observations constitute the widest fractional bandwidth and largest sky area survey at radio frequencies to date, and calibrate the low frequency flux density scale of the southern sky to better than 10\,\%. This paper presents details of the flagging, imaging, mosaicking, and source extraction/characterisation, as well as estimates of the completeness and reliability. All source measurements and images are available online\footnote{\texttt{http://www.mwatelescope.org/}}. This is the first in a series of publications describing the GLEAM survey results.

\end{abstract}


\section{Introduction}
\label{sec:introduction}

Low-frequency (50--350\,MHz) radio sky surveys are now at the forefront of modern radio astronomy, as part of preparation for the Square Kilometre Array \citep{Dewdney2015}.
Surveys are currently being performed by a range of established telescopes, such as the Giant Metrewave Radio Telescope \citep[GMRT; ][]{1991CuSc...60...95S}
and the Karl G. Jansky Very Large Array \citep[VLA; ][]{2011ApJ...739L...1P}.
The last decade has also seen the construction of new radio telescopes, the primary science driver of which is measuring the emission from high-redshift neutral hydrogen during the Epoch of Reionisation (EoR), predicted to lie between 50 and 200\,MHz \citep[e.g.][]{2006PhR...433..181F,2010ARA&A..48..127M}. These instruments include the Low Frequency Array \citep[LOFAR; ][]{2013A&A...556A...2V},
the Precision Array for Probing the Epoch of Reionization \citep[PAPER; ][]{2010AJ....139.1468P},
the Long Wavelength Array \citep[LWA; ][]{2013ITAP...61.2540E}, and the Murchison Widefield Array \citep[MWA; ][]{2009IEEEP..97.1497L,2013PASA...30....7T}.

Recent low-frequency surveys using these established and new telescopes include:
the VLA Low-frequency Sky Survey Redux at 74\,MHz \citep[VLSSr; ][]{2014MNRAS.440..327L},
the Multi-Snapshot Sky Survey at 120--180\,MHz \citep[MSSS; ][]{2015A&A...582A.123H},
the Tata Institute for Fundamental Research GMRT Sky Survey at 150\,MHz \citep[TGSS; ][]{2016arXiv160304368I}, 
and the GaLactic Extragalactic All-sky MWA (GLEAM) survey at 72--231\,MHz \citep{2015PASA...32...25W}.
These build on the success of radio sky surveys released in the late 1990s: the NRAO VLA Sky Survey \citep[NVSS; ][]{1998AJ....115.1693C} at 1.4\,GHz, the Sydney University Molonglo Sky Survey at 843\,MHz \citep[SUMSS; ][]{1999AJ....117.1578B,2003MNRAS.342.1117M}, and the Molonglo Reference Catalogue at 408\,MHz \citep[MRC;][]{1981MNRAS.194..693L,1991Obs...111...72L}.

The low observing frequencies of the new surveys increase the challenge of dealing with ionospheric distortion ($\propto\nu^{-2}$, where $\nu$ is frequency) and high sky temperatures from diffuse Galactic synchrotron emission ($\propto\nu^{-2.7}$). Low frequencies also imply larger fields-of-view, making imaging more difficult, especially when combined with position-dependent ionospheric distortion. Additionally, new aperture arrays such as LOFAR and the MWA, consisting of large arrays of mutually coupled dipoles instead of traditional parabolic dishes, present demanding challenges to calibration and imaging, with complex frequency- and spatially-variant sensitivity patterns on the sky (the ``primary beams''). The large number of antennas in these new arrays, coupled with wide fields-of-view, also gives rise to massive data volumes which are logistically and computationally challenging to handle, putting low-frequency radio astronomy on the cutting edge of Big Data problems in science.

However, the scientific return is expected to be commensurately great; detecting the EoR would constrain the last remaining unknown period in standard cosmology; measuring the low-frequency spectra of radio galaxies will open windows into high-redshift and synchrotron-self-absorbed populations; detecting the remains of steep-spectrum ancient radio jets tells us about the life cycle of active galactic nuclei; and measuring the full spectral details of relics and haloes in galaxy clusters constrains the generation mechanisms for these sources. More details of the science enabled by wide-band low-frequency observations, and specifically the MWA, can be found in \citet{2013PASA...30...31B}.

Further, from the point of view of preparing for SKA1 LOW, the precursor and pathfinder projects currently underway are deeply informing the design process for the SKA. LOFAR, the MWA, and PAPER are all exploring different configuration architectures for large-scale low-frequency arrays, different calibration strategies, and different signal extraction strategies for EoR experiments. The MWA is plays key role in this respect, being located on the site of the eventual SKA1 LOW, and exercising technical and scientific aspects of low-frequency radio astronomy in an end-to-end manner in that environment.

GLEAM is the MWA's widefield continuum imaging survey, and has surveyed the sky south of declination $+30^{\circ}$ over a frequency range of 72--231\,MHz. Much of the wisdom accumulated by the MWA team in addressing the challenges outlined above is expressed in the production of this survey. As such, the survey represents a significant step forward for a wide range of astrophysical applications and also represents substantial progress on the path to SKA1 LOW.

Here we present the first extragalactic radio source catalogue from the GLEAM survey, consisting of \nsrc sources over \survarea square degrees: the entire southern sky excluding the Magellanic Clouds and Galactic latitudes within 10\arcdeg of the Galactic Plane. A spectral resolution of 8\,MHz enables multi-frequency studies of radio galaxies, active galactic nuclei, and galaxy clusters. A drift scan imaging strategy controls systematics and reduces the number of primary beams which need to be calibrated \citep{2015PASA...32...25W}.  This paper describes the first release of survey data products and will be the first in a series of papers that will describe the GLEAM survey results over the next few years.

This paper is laid out as follows.
\Sect~\ref{sec:reduction} describes the observations, and the calibration and imaging strategy used in data reduction.
\Sect~\ref{sec:fluxscale} describes the flux density calibration procedures, including correcting for the MWA primary beam, and the estimated errors on the procedure;
\Sect~\ref{sec:sourcecat} describes the source-finding process, how both resolved and unresolved sources are characterised across the wide frequency bandwith, and the resulting sensitivity and completeness of the catalogue;
\Sect~\ref{sec:properties} describes the properties of the source catalogue such as spectral index distribution and source counts;
\Sect~\ref{sec:conclusions} contains a discussion and conclusion.

All position angles are measured from North through East (i.e. counter-clockwise).
All equatorial co-ordinates are J2000. Several figures use the ``cubehelix'' colour scheme \citep{2011BASI...39..289G}, as it is colour-blind friendly, and desaturates smoothly to greyscale.
Right ascension is abbreviated as RA, and declination is abbreviated as Dec.
The spectral index of a source, $\alpha$, is given using the convention $S_{\nu}\propto\nu^\alpha$.

\section{Observations and data reduction}
\label{sec:reduction}

\subsection{Observations}

As detailed by \citet{2013PASA...30....7T}, the MWA consists of 128 32-dipole antenna ``tiles'' distributed over an area approximately 3~km in diameter. Each tile observes two instrumental polarisations, ``X'' (16~dipoles oriented East-West) and ``Y'' (16~dipoles oriented North-South), the pointing of which is controlled by per-tile analogue beam formers.
The signals from the tiles are initially processed by 16~in-field receiver units, each of which services eight tiles.

As described in \citet{2015PASA...32...25W}, GLEAM observes in week-long drift scan campaigns, with a single Dec strip observed each night. The observing bandwidth of 72--231\,MHz is covered by shifting frequencies by 30.72\,MHz every two minutes, avoiding the Orbcomm satellite constellation at 134--139\,MHz. Thus, the frequencies of observation are 72--103\,MHz, 103--134\,MHz, 139--170\,MHz, 170--200\,MHz, and 200--231\,MHz. These may be further subdivided for imaging purposes; in this paper, the 30.72\,MHz bandwidth is commonly subdivided into four 7.68\,MHz sub-channels. The native channel resolution of these observation is 40\,kHz and the native time resolution is 0.5\,s.

This paper concerns only data collected in the first year, i.e. four weeks between June~2013 and July~2014. We also do not image every observation, since the survey is redundant across approximately 50\,\% of the observed RA ranges, and some parts are adversely affected by the Galactic plane and Centaurus~A. \Tab~\ref{tab:obs} lists the observations which have been used to create this first GLEAM catalogue.

\begin{table}
\caption{GLEAM first year observing parameters. $N_\mathrm{flag}$ is the number of flagged tiles out of the 128~available. The calibrator is used to find initial bandpass and phase corrections as described in \Sect~\ref{sec:reduction}.
\label{tab:obs}}

\begin{tabular}{ccccc}
\hline 
Date & RA range (h) & Dec ($\arcdeg$) & $N_\mathrm{flag}$ & Calibrator\tabularnewline
\hline 
\hline 
2013-08-09 & 19.5--3.5 & $-55$ & 10 & 3C444\tabularnewline
2013-08-10 & 19.5--3.5 & $-27$ & 4 & 3C444\tabularnewline
2013-08-18 & 19.5--3.5 & $-72$ & 7 & Pictor A\tabularnewline
2013-08-22 & 19.5--3.5 & $-13$ & 3 & 3C444\tabularnewline
2013-08-25 & 19.5--3.5 & $-40$ & 5 & 3C444\tabularnewline
2013-11-05 & 0--8 & $-13$ & 4 & Hydra A\tabularnewline
2013-11-06 & 0--8 & $-40$ & 4 & Hydra A\tabularnewline
2013-11-07 & 0--8 & $+2$ & 4 & Hydra A\tabularnewline
2013-11-08 & 0--8 & $-55$ & 4 & Hydra A\tabularnewline
2013-11-11 & 0--8 & $+19$ & 4 & Hydra A\tabularnewline
2013-11-12 & 0--8 & $-72$ & 4 & Hydra A\tabularnewline
2013-11-25 & 0--8 & $-27$ & 4 & Hydra A\tabularnewline
2014-03-03 & 6--16 & $-27$ & 4 & Hydra A\tabularnewline
2014-03-04 & 6--16 & $-13$ & 4 & Hydra A\tabularnewline
2014-03-06 & 6--16 & $+2$ & 4 & Hydra A\tabularnewline
2014-03-08 & 6--16 & $+19$ & 4 & Hydra A\tabularnewline
2014-03-09 & 6--16 & $-72$ & 4 & Virgo A\tabularnewline
2014-03-16 & 6--16 & $-40$ & 4 & Virgo A\tabularnewline
2014-03-17 & 6--16 & $-55$ & 4 & Hydra A\tabularnewline
2014-06-09 & 12--22 & $-27$ & 8 & 3C444\tabularnewline
2014-06-10 & 12--22 & $-40$ & 8 & 3C444\tabularnewline
2014-06-11 & 12--22 & $+2$ & 8 & Hercules A\tabularnewline
2014-06-12 & 12--18.5 & $-55$ & 8 & 3C444\tabularnewline
2014-06-13 & 12--19 & $-13$ & 8 & Centaurus A\tabularnewline
2014-06-14 & 12--22 & $-72$ & 9 & Hercules A\tabularnewline
2014-06-15 & 12--22 & $+18$ & 13 & Virgo A\tabularnewline
2014-06-16 & 18.5--22 & $-13$ & 8 & 3C444\tabularnewline
2014-06-18 & 18.5--22 & $-55$ & 8 & 3C444\tabularnewline
\hline 
\end{tabular}
\end{table}

\subsection{Flagging and averaging}\label{sec:flagging}

The raw visibility data are processed using \textsc{Cotter} \citep{2015PASA...32....8O}, which performs flagging using the \textsc{AOFlagger} algorithm \citep{2012A+A...539A..95O} and averages the visibilities to a time resolution of 4\,s and a frequency resolution of 40\,kHz, giving a decorrelation factor of less than a percent at the horizon. Approximately 10\,\% of the 40\,kHz channels are flagged due to aliasing within the polyphase filterbank. Typical flagging percentages correspond well to those found by \citet{2015PASA...32....8O}: 2--3\,\% in the frequency-modulated (FM: 80--100\,MHz) and digital TV (DTV: 190-220\,MHz) bands and 1.5\,\% elsewhere. In just $\approx0.5$\,\% of observations in the FM and DTV bands, high-intensity radio frequency interference (RFI) renders the entire 30.72\,MHz observation unuseable; these data are simply discarded, as the highly redundant observing strategy ensures relatively even sky coverage regardless.

\subsection{Calibration}\label{sec:calibration}

Snapshot calibration proceeds in three stages: an initial transfer of complex antenna-based gain solutions derived from a bright, well-modelled calibrator, a self-calibration loop, and a flux scale and astrometry correction to the resulting images. The MRC is extremely useful for calibrating both the flux density scale and astrometry of GLEAM. At 408\,MHz, the MRC is reasonably close to the GLEAM frequency range, usefully placed between VLSSr and NVSS in the Northern sky. It covers most of the GLEAM survey area, over $-85\arcdeg<\mathrm{Dec}<+18\fdg5$, and contains 12,141 discrete sources with $S>0.7$\,Jy, 92\% of which are isolated and point-like ($\mathrm{MFLAG}=0$), at a resolution of $2\farcm62$ by $2.86\sec(\mathrm{Dec}+35\fdg5)\arcmin$. We make use of this catalogue frequently during calibration, and when we set the final flux scale of the catalogue (\Sect~\ref{sec:fluxscale}).

\subsubsection{Initial calibration}\label{sec:initcal}

Electrical delays have been applied to the instrument such that less than 180\,$^\circ$ of phase rotation is evident over the observing bandwidth of 30.72\,MHz. Given that GLEAM mostly consists of sky previously unobserved at these frequencies and angular scales, a typical observation cannot be immediately self-calibrated using an existing model of the field. We thus perform an initial calibration in a similar manner to \citet{2014PASA...31...45H}: for each of the five frequency bands, we observe a specific bright calibrator source (\Tab~\ref{tab:calibrators}), and Fourier-Transform a model of the source derived from other low-frequency measurements, scaled by a model of the primary beam \citep{2015RaSc...50...52S} to create a model set of visibilities. Per-40kHz-channel, per-polarisation, per-antenna complex gains are created using a least-squares fit of the data to the model via the Common Astronomy Software Applications (\textsc{CASA}\footnote{http://casa.nrao.edu/}) task \textsc{bandpass}.
In typical extra-Galactic sky, baselines shorter than 60\,m are not used to perform calibration, due to contamination from expected large-scale Galactic emission. The calculated antenna amplitude and phase solutions are then applied to the entire night of drift scan data.

\Tab~\ref{tab:calibrators} shows the calibrators used for each drift scan, including their approximate 200-MHz flux densities and spectral indices. The aim of this initial calibration is to bring the flux scale to within $\approx20$\,\% of the literature values for typical sources, and the phases to within $\approx5^\circ$ of their correct values. This allows the creation of an initial sky model from each observation, in order to begin a self-calibration loop.

\begin{table*}
\caption{Sources used for initial bandpass calibration and/or peeled from the data,
with approximate flux densities and spectral indices calculated using measurements over 60--1400\,MHz available via the NASA/IPAC Extragalactic Database (NED)\protect\footnotemark. Exact flux densities are not needed because the data are self-calibrated during peeling (\Sect~\ref{sec:peeling}), and every observation is later rescaled to a single flux scale (see \Sects~\ref{sec:imaging} and \ref{sec:corr_pb}).\label{tab:calibrators}}

\begin{tabular}{cccccc}
\hline 
Source & RA & Dec & $S_{200\mathrm{MHz}}$/ Jy & $\alpha$ & \textbf{C}alibrator/\textbf{P}eeled\tabularnewline
\hline 
\hline 
3C444 & 22 14 26 & $-17$ 01 36 & 60 & $-0.96$ & C\tabularnewline
Centaurus A & 13 25 28 & $-43$ 01 09 & 1370 & $-0.50$ & C\tabularnewline
Hydra A & 09 18 06 & $-12$ 05 44 & 280 & $-0.96$ & C, P\tabularnewline
Pictor A & 05 19 50 & $-45$ 46 44 & 390 & $-0.99$ & C, P\tabularnewline
Hercules A & 16 51 08 & $+04$ 59 33 & 377 & $-1.07$ & C, P\tabularnewline
Virgo A & 12 30 49 & $+12$ 23 28 & 861 & $-0.86$ & C, P\tabularnewline
Crab & 05 34 32 & $+22$ 00 52 & 1340 & $-0.22$ & P\tabularnewline
Cygnus A & 19 59 28 & $+40$ 44 02 & 7920 & $-0.78$ & P\tabularnewline
Cassiopeia A & 23 23 28 & $+58$ 48 42 & 11900 & $-0.41$ & P\tabularnewline
\hline 
\end{tabular}
\end{table*}

\subsubsection{Peeling}\label{sec:peeling}
As the MWA dipoles are arranged in a regular grid in every tile, and the field-of-view is wide, sources also appear in primary beam grating sidelobes with sensitivity levels of order 10\,\% of the main beam response. These sidelobes lie 45--$90^\circ$ from the main lobe: imaging them for every observation would be prohibitively expensive. Typically, the sidelobes observe only ``faint'' extra-Galactic sky, but occasionally a bright ($>100$\,Jy) source may cause significant signal in the visibilities. In these cases, further processing is made considerably easier by first ``peeling'' such a source from the visibilities. An automatic script reads a list of bright sources' positions and flux densities and multiplies it by a model of the primary beam for that observation. Whenever a source is expected to be more than 50\,Jy in apparent flux density, it is peeled from the visibilities. Sources fainter than this do not upset the self-calibration loop.

In a similar fashion to the initial calibration step (\Sect~\ref{sec:initcal}), a model of the source is multiplied by the model of the primary beam, and Fourier-Transformed into visibility space. A self-calibration is performed (in both amplitude and phase), the gains applied to these model visibilities and then the source is subtracted from the main visibility dataset. We do not apply the derived gains to the main visibility dataset, because the primary beam grating lobes have steep spectral behaviour, which would cause an unusual amplitude gain factor; the sky covered by the sidelobes also experiences different ionospheric conditions to the main lobe, so applying these gains would cause a phase distortion in the main lobe. \Tab~\ref{tab:calibrators} shows the sources peeled from the GLEAM data.

\footnotetext{http://ned.ipac.caltech.edu/}
\subsubsection{Imaging and self-calibration}\label{sec:imaging}

The widefield imager \textsc{WSClean} \citep{2014MNRAS.444..606O} is used for all imaging, as it deals with the wide-field $w$-term effects using $w$-stacking, can produce a useful projection for our data, and performs deep imaging in a reasonable time ($\approx3$\,hours per 2-minute observation) in a multi-threaded fashion, suitable for use on supercomputers. All imaging is performed on a per-observation basis, forming 2-minute ``snapshots'' of the sky.

Throughout the imaging process, we image the primary beam down to the 10\,\% level, corresponding to squares of 4000\,pixels on each side. Pixel scales are chosen such that the width at half its maximum value (full-width half-maximum; FWHM) of the synthesised beam is always sampled by at least four pixels. For example, at the lowest frequency of 72\,MHz, the pixel size is $57\farcs3\times57\farcs3$, and the imaged field-of-view is $63\fdg8\times63\fdg8$; for the highest frequency of 231\,MHz, the pixel size is $23\farcs4\times23\farcs4$ and the imaged field-of-view is $26\fdg0\times26\fdg0$.

We use an optimum $(u,v)$-weighting scheme for imaging compact objects with the MWA: the ``Briggs'' scheme, using a ``robust'' parameter of $-1.0$ (close to uniform weighting) \citep{1995AAS...18711202B}. All linear instrumental polarisations are imaged, and during the \textsc{clean} process, peaks are detected in the summed combination of the polarisations, but components are refitted to each polarisation once a peak location is selected. The projection used is a SIN projection centred on the minimum-$w$ pointing, i.e. hour angle~$=0$, Dec~$-26.7^\circ$. The restoring beam is a 2-D Gaussian fit to the central part of the dirty beam, and remains very similar in shape (within 10\,\%) for each frequency band of the entire survey, which will be important later for mosaicking (\Sect~\ref{sec:mosaicking}).

Using \textsc{chgcentre}, part of the \textsc{WSClean} package, the peeled visibilities are phase-rotated to the minimum $w$-term direction, within $1^\circ$ of zenith. This optimises the speed and memory use of the $w$-stacking algorithm \citep{2014MNRAS.444..606O}. The observation is imaged across the entire 30\,MHz bandwidth using multi-frequency synthesis, in instrumental polarisations (XX, XY, YX, YY), down to the first negative clean component, without any major cycles. Given the noise-reducing effects of the initial calibration and peeling steps described above, this typically results in models of total flux density $\approx50-200$\,Jy, with around one clean component per square degree. (For comparison, final models usually have $\approx25$~clean components per square degree and a total flux density $\approx100-400$\,Jy.)

These instrumental polarisation images are combined using the complex primary beam \citep{2015RaSc...50...52S} into astronomical Stokes (I, Q, U, V). Based on previous polarisation observations with the MWA 32-tile prototype \citep{2013ApJ...771..105B}, we expect the vast majority of clean components to be unpolarised, so set Q, U and V to zero, in order that the initial self-calibration model is purely unpolarised. The Stokes~I image is transformed using the same primary beam model back into instrumental polarisations, and this new model is Fourier-transformed to create a set of model visibilities, which are used to derive new per-40-kHz-channel per-polarisation per-antenna complex gain solutions (in both amplitude and phase), over the whole observing interval of 112\,s.
Identically to the initial calibration step, baselines shorter than 60\,m are not used to determine calibration solutions.

The new gains are then applied to the visibilities. \textsc{AOFlagger} is rerun on the visibilities in order to flag any RFI which was missed by the initial flagging step (\Sect~\ref{sec:flagging}), and is easier to detect now that the data are calibrated. This is particularly useful for the FM and DTV bands, where typically another $\approx1$\,\% of the bandwidth is flagged per observation. As a measure of image noise, the root-mean-squared (RMS) of the initial image is measured and a new \textsc{clean} threshold is set to three times that RMS: typical values of this new \textsc{clean} threshold are 240--60\,mJy from 72--231\,MHz.

\begin{figure}
	\centering
	\includegraphics[width=0.5\textwidth]{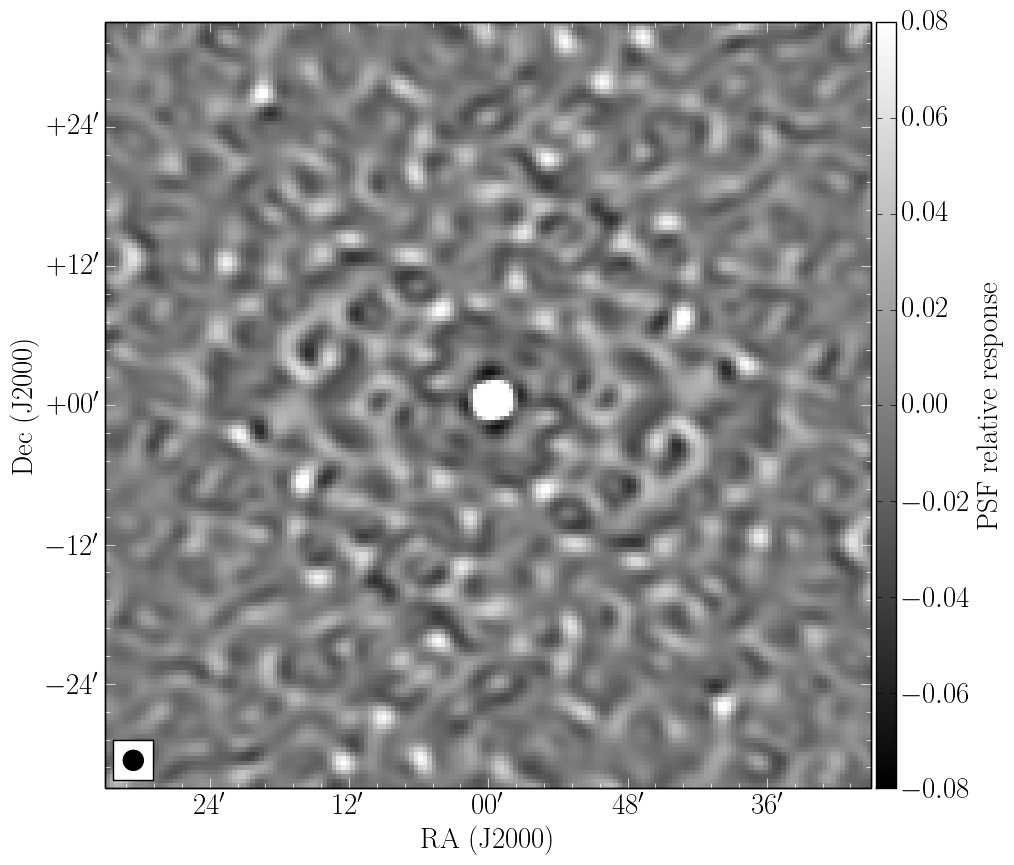}
	\caption{Central square degree of the synthesised beam of the MWA at 200--208\,MHz, showing the low sidelobe levels.}
    \label{fig:PSF}
\end{figure}

At this stage, we divide the 30.72\,MHz bandwidth of each observation into narrower sub-bands of 7.68\,MHz, which will be used throughout the rest of the paper. This sub-band width was chosen as a compromise between various factors. As the width of a sub-band increases, the synthesised beam sidelobes are reduced, and this alongside higher sensitivity enables deeper deconvolution. However, the primary beam model, generated at the centre of the sub-band, becomes less correct for the edges of the sub-band, as the bandwidth increases. The logistics of processing a large amount of data becomes more difficult depending on how much the data is sub-divided, while the potential usefulness of the catalogue increases with higher frequency resolution, as long as sensitivity is not overly reduced. Dividing the 30.72\,MHz bandwidth into four 7.68-MHz sub-bands is a good compromise between these competing factors. The central area of the synthesised beam of a 7.68-MHz sub-band at 200\,MHz is shown in \Fig~\ref{fig:PSF}, showing minima and maxima of order 8\,\% of the peak response.

Using \textsc{WSClean}, we jointly clean these sub-bands, also jointly searching for \textsc{clean} components across the instrumental polarisations, as in the initial imaging step. Major cycles are performed: typically four per observation. When the \textsc{clean} threshold is reached, the \textsc{clean} components are restored to make the instrumental polarisation images. These are again transformed using the primary beam model to make astronomical Stokes images.

MRC is then used to set a basic flux scale for the snapshot images. Typically the self-calibrated images have a flux scale 10--20\% lower than the initial images, as only 80--90\% of sky flux is captured in the initial model at the start of the self-calibration loop. We note that the loop does not cause sources uncaptured in the initial model to become even fainter, as they would in a telescope with fewer antennas (self-calibration bias); instead, all sources become fainter. We also note that while it would be ideal to perform snapshot calibration on a scaled sky model, from e.g. VLSSr and MRC, attempts to do so resulted in lower-fidelity images than a direct self-calibration, likely due to poorly-extrapolated source spectral indices and morphologies. Instead, we rescale the flux scale by selecting a sample of sources and cross-matching them with MRC, then compare the measured flux densities with those predicted from MRC. Failing to do this would lead to flux scale variations of order 10--20\% between snapshots.

For the purposes of per-snapshot flux calibration, source-finding is performed using \textsc{Aegean}
v1.9.6 \citep{2012MNRAS.422.1812H} \footnote{https://github.com/PaulHancock/Aegean}
on the primary-beam-corrected Stokes~I images, using a minimum threshold of $8\sigma$ (typically 1.6--0.3\,Jy). Unresolved sources are selected by using only sources where the integrated flux density is less than twice the peak flux density. Sources in positions where the primary beam has $<20$\,\% of the maximum primary beam sensitivity are discarded, as are any peeled and restored bright ($S>100$\,Jy) sources (\Tab~\ref{tab:calibrators}). The snapshot catalogue and MRC are cross-matched using the Starlink Tables Infrastructure Library Toolset \citep[STILTS;][]{2006ASPC..351..666T}, using only MRC sources identified as morphologically simple, and isolated ($\mathrm{MFLAG}=0$). The MRC flux densities are scaled to the relevant snapshot frequency using an assumed spectral index $\alpha = -0.85$. (Note that the precise flux scale is irrelevant as a more thorough flux calibration is performed later; see \Sect~\ref{sec:fluxscale}). We calculate the ratio of the scaled MRC flux densities and the measured snapshot integrated flux densities, weighting by the square of the signal-to-noise (S/N) of the sources in the GLEAM snapshot, and use this ratio to scale the snapshot. Typically around 250~sources are used to perform the flux calibration. This removes any remaining RA- (time-) dependent flux scale errors from the drift scans; these are typically of order 10--20\,\%.

Typical expected snapshot RMS values are 200\,mJy at 72\,MHz to 40\,mJy at 231\,MHz. After the rescaling, the RMS of the image is measured, and if it is more than double the expected value for the band, the snapshot is discarded. This removes a further $\approx2$\,\% of snapshots.

\subsection{Astrometric calibration}\label{sec:astrometry}

Per-snapshot position offsets are introduced from ionospheric distortions, which vary slowly over the night. There is usually also a small ($<20\arcsec$) astrometry error constant across all snapshots that is introduced at the initial calibration stage, for two reasons: the calibrator models used are generally scaled from high-frequency observations, where the morphology of the source may be different, and the calibrator observations are often made at a different location in the sky, resulting in a different refractive angle through the ionosphere.


For Declinations south of $18\fdg5$, the MRC catalogue is used to determine the reference positions of sources. North of this Dec, we form a similar catalogue, by cross-matching NVSS and VLSSr, and calculating a 408\,MHz flux density assuming a simple power law spectral index for every source ($S\propto\nu^\alpha$), and then discarding all sources with $S_\mathrm{408\,MHz}<0.67$\,Jy, the same minimum flux density as MRC. For these remaining sources, the NVSS positions are used as the reference source positions. For sources detected in each GLEAM snapshot, we use the same source size and signal-to-noise filters as the previous section. We crossmatch the ``extended" MRC catalogue with every GLEAM snapshot catalogue, and calculate the position offsets for every source. In each snapshot there are 100--1000 crossmatched sources, depending on observation quality and frequency.

\begin{figure*}
\centering
\includegraphics[width=\textwidth]{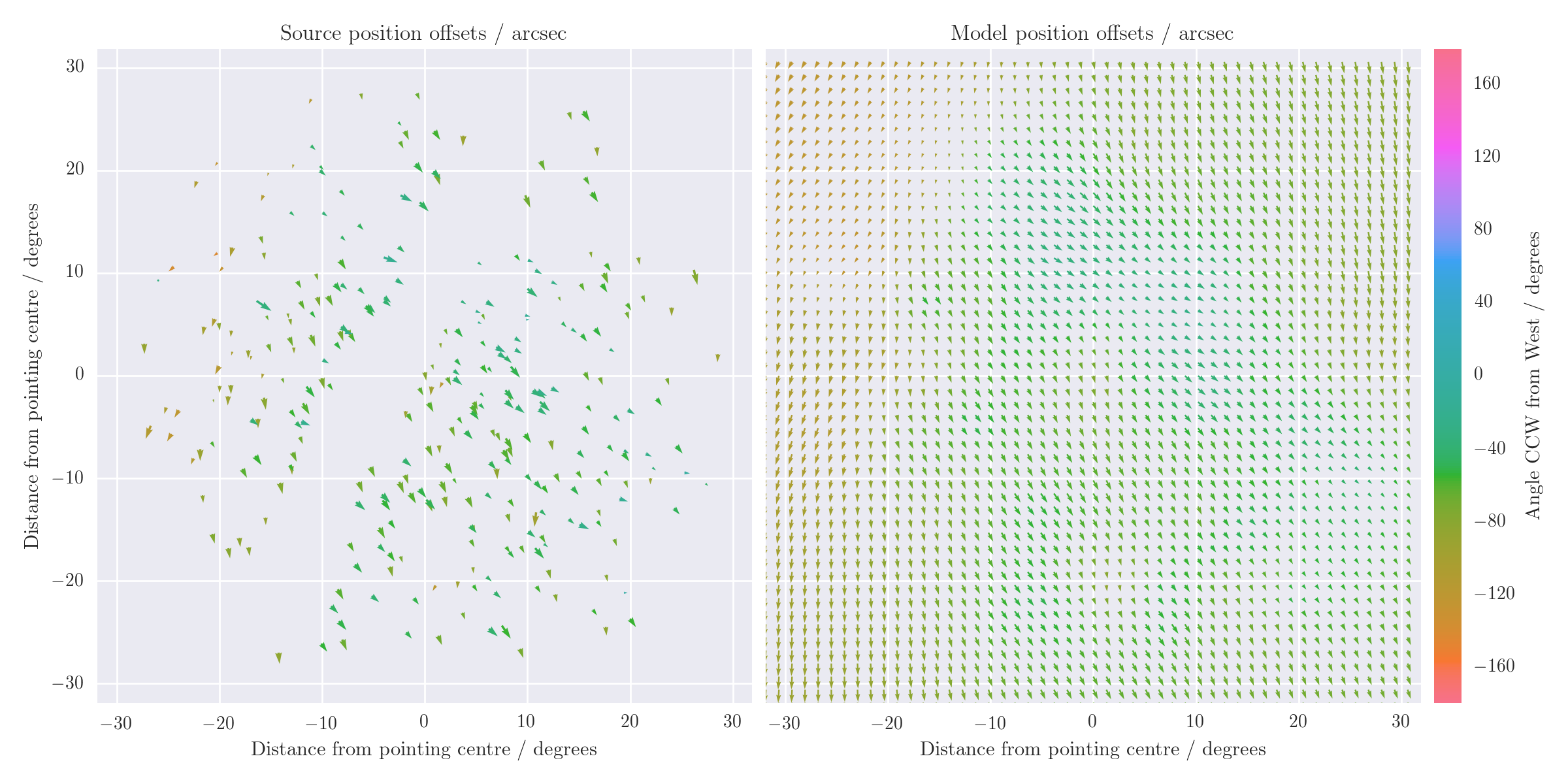} 
    \caption{An animation, at four frames per second, of the measured and modelled ionospheric distortions for 44 observations performed on the night of 2013~Nov~25, at 72--80\,MHz, where the distortions are largest. The left panel shows source position offsets, as measured by comparing the positions of 8-$\sigma$ unresolved sources with the MRC and NVSS catalogues (see \Sect~\ref{sec:astrometry} for a full description of the cross-matching). The vectors indicate the direction of the correction that needs to be applied to align GLEAM with the reference catalogues. The right panel shows a radial basis function fit to these vectors, with values shown over a grid of $50\times50$ points. In both panels, the axes are plotted in $(l,m)$ rather than in equatorial co-ordinates. The diagonal gap in the source offset measurements in the last ten observations is caused by the Galactic Plane. In printed versions, only the last frame will display.}
    \label{fig:vector-movie}
\end{figure*}

Equatorial celestial co-ordinates are not the correct reference frame for determining widefield ionospheric corrections, particularly near the South Celestial Pole. Therefore, we convert our RA and Dec offsets into $(l,m)$ offsets. We fit a radial basis function to these offsets, with a scale size of $10\arcdeg$, the typical scale size of ionospheric distortions. \Fig~\ref{fig:vector-movie} shows the raw source position offsets and the fitted radial basis function, for a single night of observing. The radial basis function is then applied to the original snapshot image, and the original pixel data interpolated onto the modified grid using a Clough-Tocher 2D interpolator \citep{CloughTocher65}. The divergence of the ionospheric offset vectors is typically very low, much less than 0.1\,\%, so no flux correction needs to be made to the resulting image data.

After astrometric correction, the per-snapshot difference RMS is of order 10--15\arcsec~ at the lowest frequencies and 3--6\arcsec~ at the higher frequencies. Based on the FWHM of the synthesised beam (210--93\arcsec) and the S/N of the sources used (typically 20), one would expect the RMS from measurement error to be of order 5\arcsec~ at the lowest frequencies and 2\farcs5~ at the highest frequencies \citep{1999ASPC..180..301F}. We conclude that there is a residual direction-dependent ionospheric distortion of magnitude 5--10\arcsec~ at the lowest frequencies, and 0.5--2.5\arcsec~ at the higher frequencies. While a visibility-based direction-dependent calibration for each snapshot would be ideal, at the time of writing, the algorithms and computational resources were not available to perform this over the entire surveyed area.


%
\begin{figure*}
\centering
\includegraphics[width=\textwidth]{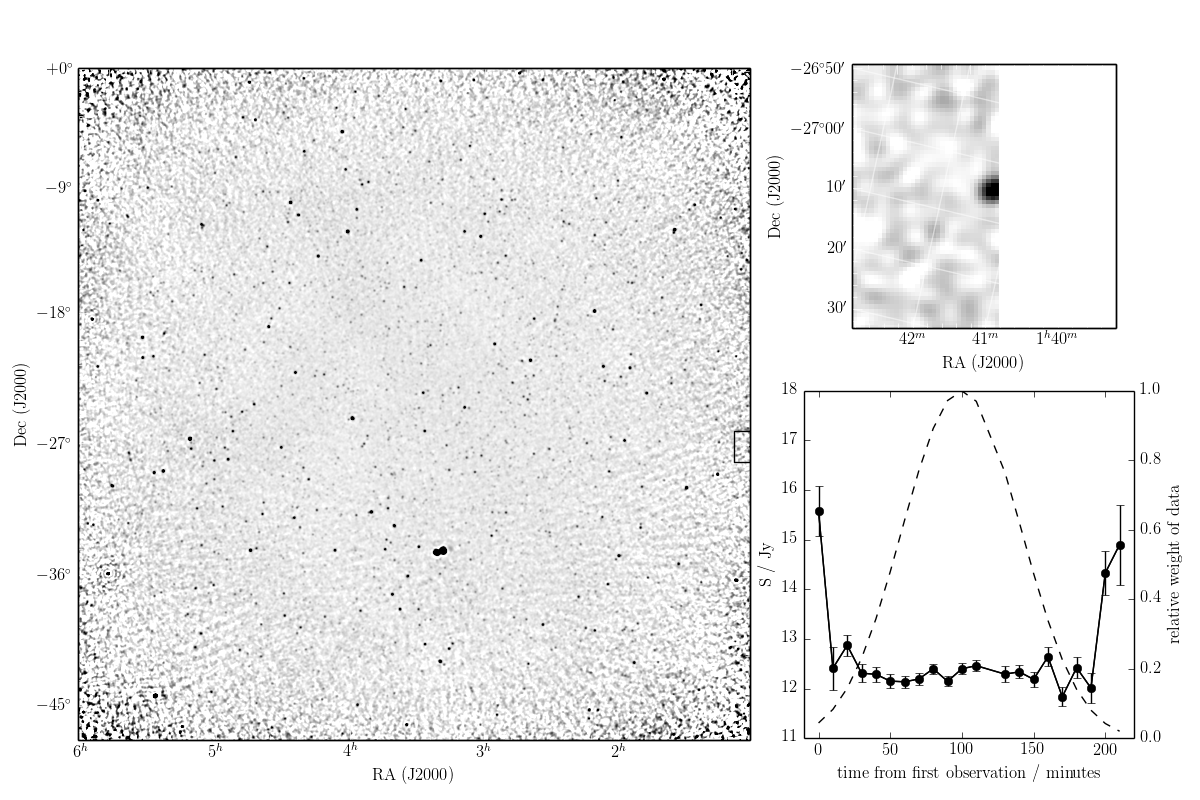} 
    \caption{The left panel shows an animation, at four frames per second, of the central $45^\circ\times45^\circ$ of the Dec\,$-27$ 103--111\,MHz drift data from the first twenty observations taken on the night of 2013~Nov~25, and following the imaging procedure outlined in \sect~\ref{sec:imaging}.
The central Dec remains constant throughout at Dec$=-26^{\circ}47\arcmin$; the first frame is centred on RA$=00^{\mathrm{h}}02^{\mathrm{m}}$ and the last is centred on RA$=03^{\mathrm{h}}23^{\mathrm{m}}$. The greyscale is linear and runs from $-0.1$--1\,Jy\,beam$^{-1}$, and a primary beam correction has been made to produce a pseudo-Stokes-I view (not truly Stokes~I, as the polarisation cross-terms are discarded, as described in \Sect~\ref{sec:mosaicking}). In order to enhance the visibility of the point sources in this figure, the images in this panel have been convolved with a Gaussian of FWHM~5\arcmin. A square indicates the area shown in the top right panel, which shows PKS~J0141-2706 and the surrounding $45\arcmin\times45\arcmin$, with a linear grayscale from $-1$--10\,Jy\,beam$^{-1}$. The bottom right panel shows the measured integrated flux density of PKS~J0141-2706 in each image (points; solid lines), with the local RMS of the image shown as an error bar. The dashed line shows the square of the primary beam response, which, combined with a global image RMS measurement, is the weighting given to each measurement at the mosaicking stage. NB: Due to size limits, the animation is only visible in the online version of this article. In printed versions, the last frame will display; note that in this frame, the test source is on the edge of the image, so while it is detected, its measured flux density is unreliable.}
    \label{fig:drift-movie}
\end{figure*}

\subsection{Mosaicking}\label{sec:mosaicking}

After flagging, self-calibration, imaging, basic flux calibration and bulk ionospheric correction have been performed, there exist, for each night, of order 3,200 snapshots of the sky, in the four instrumental polarisation parameters and four 7.68-MHz frequency bands for each observation. As an example, the first twenty primary-beam-corrected and flux-calibrated snapshots of the 103--111\,MHz observations of the night of 2013~Nov~25 are shown in \Fig~\ref{fig:drift-movie}, as well as a randomly-chosen typical $\approx12$\,Jy source, PKS~J0141-2706, and its integrated flux density as measured in each snapshot. As is particularly visible at the edges of the image, correcting the snapshots to astronomical Stokes using the current primary beam model does not result in flat flux calibration across the image compared to literature values: there are residual errors in the model of order 5--20\,\%, worst at frequencies $>180$\,MHz, at distances of more than $20^\circ$ from the centre of the primary beam, and at zenith angles of more than $30^\circ$. For this reason we do not combine the XX and YY snapshots (or perform any averaging across frequency bands) at this stage, and instead follow a similar method to \citet{2014PASA...31...45H} to separately flux-calibrate these polarisations before combining them into pseudo-Stokes~I. As there are complex errors in the primary beam model, but compact objects are not typically strongly polarised at low frequencies, we are unable to correct the instrumental cross-terms (XY, YX); these are discarded at this stage.

Throughout, we use the mosaicking software \textsc{swarp} \citep{2002ASPC..281..228B}. To minimise flux loss from resampling, images are oversampled by a factor of four when regridded, before being downsampled back to their original resolution.
When generating mosaics, we weight each snapshot by the square of its primary beam response; although we know the primary beam model to be inaccurate, we do not have the S/N to derive a new primary beam model based on the snapshots alone, particularly at the fainter edges toward the null. This weighting is shown as a dashed line in the bottom right panel of \Fig~\ref{fig:drift-movie}. The strong weighting ensures that where the beam model is most inaccurate ($<50$\,\% of the full response), it is most strongly downweighted (by a factor of $>4$ moving outward from that half-power point). We also include an inverse-variance weighting based on the typical RMS of the centre of each snapshot, which optimises the mosaics toward better S/N, downweighting any snapshot with residual sidelobes from poorly-peeled sidelobe sources or residual RFI.

In essence, this follows equation (1) from \citet{1996A+AS..120..375S} which maximises the S/N in the output mosaic given the changing S/N over the field in each snapshot due to the primary beam.

For each night, the mosaicking process forms 40\,mosaics from the $20\times7.68$-MHz sub-bands and the two polarisations, XX and YY.

\section{Flux scaling and primary beam corrections}
\label{sec:fluxscale}

\subsection{Matching the polarisations}
\begin{figure*}
	\centering
	\includegraphics[width=0.8\textwidth]{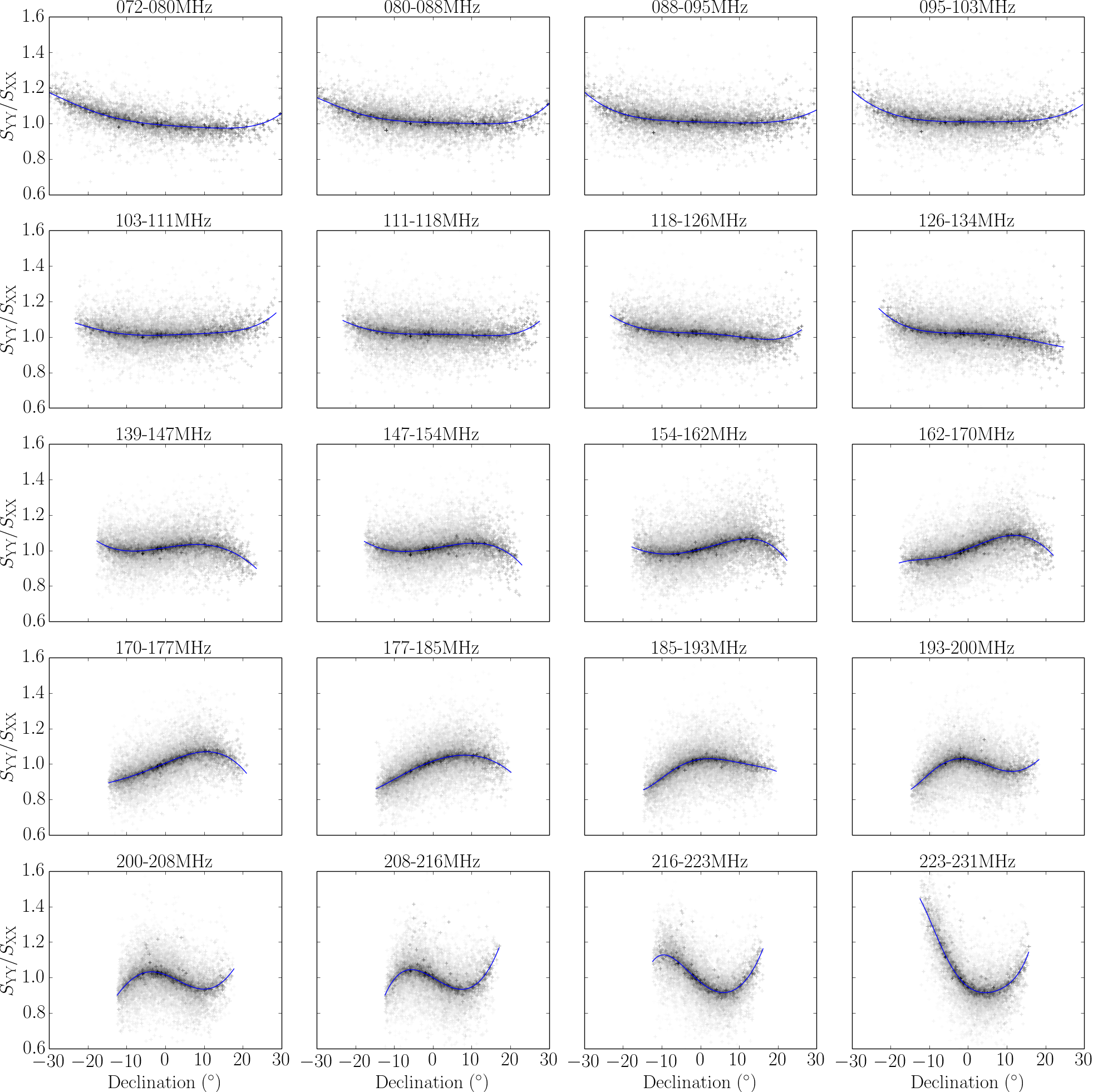}
	\caption{Ratios of source peak fluxes measured in YY and XX mosaics from 2013-11-07, centered on Dec~$+2^\circ$, with respect to Dec. All 20 sub-bands are shown, from lowest (top-left) to highest (bottom-right) frequencies. The weight of the data is the quadrature sum of the S/N of the sources in each mosaic, and is represented in a log grayscale, with darker points having higher weight. Fifth-order polynomial fits to the data are shown as blue solid lines. The Dec range sampled reduces as the primary beam becomes narrower with increasing frequency.}
    \label{fig:XX-YY}
\end{figure*}

The first step we take is to rescale the XX mosaics to match the YY mosaics, for which the beam model is slightly more accurate, as the YY (E--W) dipole is not foreshortened in a meridian drift scan. This is carried out by source-finding on each mosaic, setting a minimum threshold of $8\sigma$ and excluding all resolved and peeled sources. The two catalogues are cross-matched, and a fifth-order polynomial is fit to the ratio of the flux densities of the sources measured in the two polarisations, with respect to Dec. This polynomial is applied to the XX mosaic to rescale it to match the YY mosaic. A detailed RMS map of each polarisation is formed using the Background and Noise Estimator (\textsc{BANE}) from the \textsc{Aegean} package, and used as the input to an inverse-variance weighted addition of the two mosaics. This forms a pseudo-Stokes-I mosaic for each sub-band, which is used from here onward for further flux calibration. \Fig~\ref{fig:XX-YY} shows the typical ratios and calculated polynomial corrections for a night of GLEAM observations.

\subsection{Correcting the primary beam error and establishing the flux density scale}
\label{sec:corr_pb}

From empirical measurements of the primary beam it was discovered that the analytical primary beam model is incorrect on the order of $\sim$5\,\% to $\sim$20\,\% \citep{2015RaSc...50...52S}, with the magnitude depending on the distance from the pointing centre and the observed Dec strip. While such primary beam uncertainties are minimised in the mosaicking procedure of the snapshots, there is still an observed residual variation in the flux density with respect to Dec in the mosaics. To correct for this Dec-dependent variation in flux density, and in the process set the flux density scale empirically, we fit polynomial functions to the ratio of the measured and predicted flux densities of bright sources, with respect to Dec, then apply the calculated correction factors.

While the primary beam model is also dependent on Dec, polynomials do not need to be derived for every Dec strip because GLEAM only uses four unique analogue beamformer settings to perform the drift scans.
This means that three of the settings are mirror-images of each other: Dec\,$-40\arcdeg$ and Dec\,$-13\arcdeg$, Dec\,$-55\arcdeg$ and Dec\,$+2\arcdeg$, and Dec\,$-72\arcdeg$ and Dec\,$+19\arcdeg$.
Therefore, a Dec-dependent primary beam correction derived for any of these pointings can be applied to its mirror Dec strip by transposing and reversing the correction with respect to zenith.
(The zenith pointing at Dec\,$-27\arcdeg$ is of course symmetric about zenith itself.)
In order to interpolate predicted spectra over the GLEAM band, rather than extrapolating down in frequency, we require a measured flux density at or below our lowest frequency measurement of 72--80\,MHz.
Fortunately, we can use VLSSr at 74\,MHz, but only for Decs greater than $-30\arcdeg$.
Therefore, we derive polynomial corrections for Dec\,$-13\arcdeg$, Dec\,$+2\arcdeg$ and Dec\,$+19\arcdeg$, and apply those corrections to both the original mosaics, and Dec\,$-40\arcdeg$, Dec\,$-55\arcdeg$, and Dec $-72\arcdeg$, respectively.
Since the primary beam is dependent on frequency, polynomials are independently derived for each of the twenty 7.68\,MHz frequency bands. The calculation proceeds as follows.

Unresolved sources 8$\sigma$ above the RMS noise floor of an individual 7.68\,MHz MWA mosaic are identified and crossmatched with the VLSSr (74\,MHz), MRC (408\,MHz), and NVSS (1400\,MHz). Each source is required to have an unresolved counterpart in all three catalogues and a VLSSr counterpart that has a flux density $>$2\,Jy, so as to minimise the influence of any systematic biases present in VLSSr. Additionally, the source must be more than two degrees away from the edge of the mosaic, have an absolute Galactic latitude greater than 10\arcdeg, remain unresolved at all frequencies, and be flagged as isolated in MRC ($\mathrm{MFLAG}=0$) and NVSS. The spectral energy distributions (SEDs) of these sources formed from the VLSSr, MRC and NVSS flux density measurements must also be well fit by a power law ($\chi^{2} < 5$: for the two degrees of freedom, $p < 0.08$) and have a spectral index $\alpha$ less than $-0.5$. This excludes sources that are doubles at the resolution of NVSS, and sources that have spectral curvature or a flat spectrum, which are more likely to be variable. The resulting average source density is approximately three sources per square degree.

Correction factors are derived by comparing the measured GLEAM flux density and the expected flux density derived from the power law fit to the VLSSr, MRC and NVSS flux density measurements. A polynomial is then fit to the correction factors as a function of Dec, weighted by the square of the S/N of the source. A cubic polynomial was always favoured over other polynomial orders as assessed by the Bayesian fitting procedure described in \S\,3 of \citet{Callingham2015}. For zenith, only sources with Dec$>-26\fdg7$ were fit.

The corrections are applied to the original mosaics and also mirrored across the zenith and applied to the corresponding mosaic on the opposite side of the sky.
Note that while the corrections are symmetric in \textit{elevation}, the mosaics are formed in RA and Dec. The sky rotates more slowly through the primary beam with increasing $|\mathrm{Dec}|$ away from the equator, leading to a different amount of time on source. This effect is calculated and removed by measuring and applying the ratio of the XX:YY corrections North and South of the zenith.

\begin{figure*}
	\centering
	\includegraphics[width=0.8\textwidth]{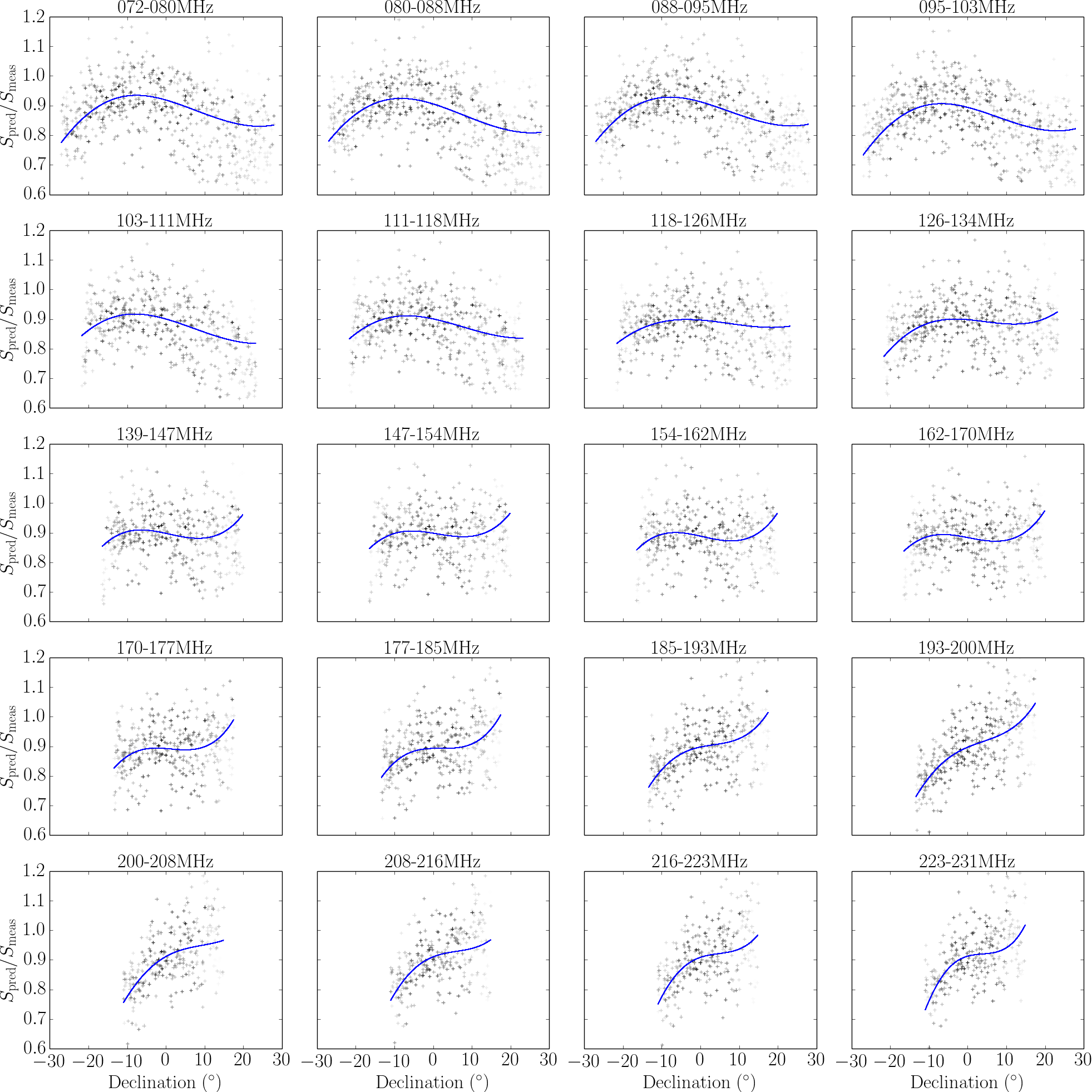}
	\caption{The ratio of the predicted flux density from the SEDs formed from VLSSr, MRC and NVSS flux density measurements, to the GLEAM flux density measurements, for the Dec\,$+2\arcdeg$ strip, from observations taken on the night of 2013-11-07. All twenty subband frequencies of GLEAM are presented, with obvious deviations from unity largest for the highest frequencies. The weight of the data is the quadrature sum of the S/N of the sources in the mosaic, and is represented in a log grayscale, with darker points having higher weight. Third-order polynomial fits to the data are shown as blue solid lines. The Dec range sampled reduces as the primary beam becomes narrower with increasing frequency.}
    \label{fig:flux_scale_dec-13}
\end{figure*}

\begin{figure*}
	\centering
	\includegraphics[scale=0.5]{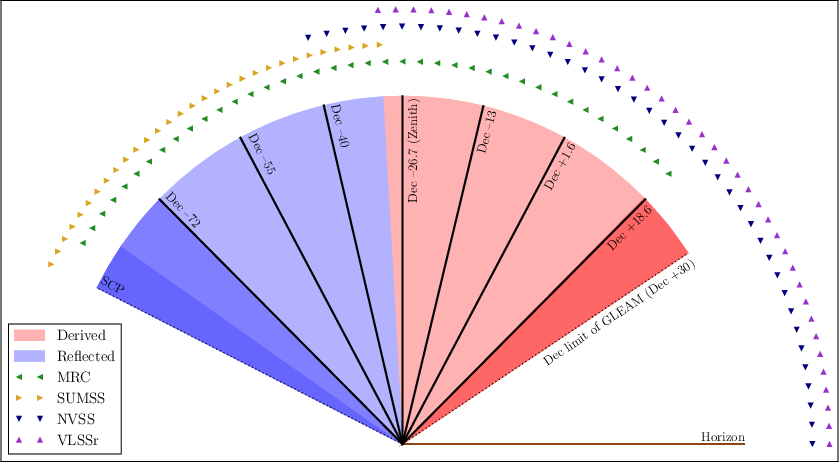} 
	\caption{A schematic demonstrating the Dec coverage of GLEAM relative to other radio surveys and the correction procedures performed for the different drift scans. The Dec settings of the drift scans performed are plotted as solid black lines. The Dec coverage of VLSSr, MRC, SUMSS and NVSS, relative to the GLEAM survey, are represented by coloured symbols. Whether the correction of the variation in flux density with Dec was derived or reflected from a mirror Dec setting is conveyed by red and blue shading, respectively. The darker red or blue shading represents greater uncertainty in the flux density scale in that declination range, as discussed in Section\,\ref{sec:corr_pb}. Note that in the final images, the transitions are not sharp as depicted here, as the data are inverse-variance weighted by the primary beam before co-addition.}
    \label{fig:fan_diagram}
\end{figure*}

An example of the polynomials derived for each frequency at Dec\,$+2\arcdeg$ is provided in \Fig~\ref{fig:flux_scale_dec-13}.
A schematic of the different correction procedures, and how the GLEAM survey compares to Dec coverage to other well-known radio surveys, is presented in \Fig~\ref{fig:fan_diagram}.

\begin{figure}
	\centering
	\includegraphics[width=0.5\textwidth]{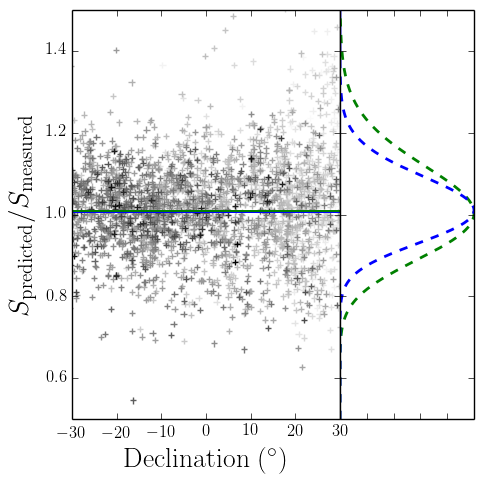}
	\caption{The overall flux scale of the 151\,MHz narrow-band image, compared to source models extrapolated from VLSSr, MRC and NVSS, as described in \Sect~\ref{sec:corr_pb}. The ordinate axis shows the ratio of the predicted to measured flux densities and the abscissa shows Dec. The greyscale of the points is the log of their weights, given by their S/N in the wideband image. The curves on the right show weighted log-Gaussian fits to the distribution of the ratios: the outer green dashed line shows the fit for those points with $\mathrm{Dec}>18\fdg5$, for which no MRC data was available, while the blue dashed line shows the fit for those points with $\mathrm{Dec}<18\fdg5$, covered by MRC. The standard deviation of the high-Dec curve is 13\,\% and the standard deviation of the low-Dec curve is 8\,\%. The plots and statistics for other narrow bands are almost identical.}
    \label{fig:correction_survey}
\end{figure}

Exploiting the symmetry of the primary beam over the meridian in correcting the flux density Dec-dependence is contingent on the flux density variation being solely due to inaccuracies in the primary beam model. It is possible the ionosphere could produce a similar Dec-dependence due to variation in electron column density with respect to elevation, and variations of ionospheric conditions each observing night \citep[see e.g.][]{2015RaSc...50..574L}. However, we are confident that the observed Dec-dependence is primarily due to deficiencies in the primary beam model because the polynomials derived for the same section of the sky on different observing nights were found to be identical. Additionally, each mosaic of a Dec strip has overlap regions with the bracketing Dec strips taken three months apart. The polynomials derived independently in the bracketing mosaics are identical in the overlap regions, further suggesting the Dec-dependence in the flux density is due to the primary beam model. This is demonstrated in \Fig~\ref{fig:correction_survey}, which shows the Dec dependence in the correction factors is flat after the polynomials have been applied across the whole survey above a Dec of $-30\arcdeg$. 

The Dec-dependent correction procedure alters slightly at Dec~$>+18\fdg5$ as this is the northern limit of MRC. Hence, sources that have a Dec greater than $18\fdg5$ have only two flux density measurements from which to estimate the correct flux density at the GLEAM observing frequency. This substantially increases the spread in correction factors due to the increase in contamination by sources with spectral curvature. Therefore, the precision of the Dec-dependent flux scale correction is lower for sources with Decs greater than $+18\fdg5$ and less than $-72\arcdeg$, the latter due to the mirroring of the corrections. This is conveyed by increased systematic uncertainties in the flux density for sources above and below these Decs (see \Sect~\ref{sec:systematic-error}).

Note that in this correction process the flux density measurements in VLSSr are converted from the flux density scale of \citet[RCB;][]{1973AJ.....78.1030R} to the flux density scale of \citet{Baars1977} for this analysis. The Baars flux density scale is less accurate than the RCB flux density scale below 300\,MHz \citep{1990MNRAS.243..637R} but such inaccuracy is on the order of $\sim$3\,\% or less (Perley et al., in prep.), and smaller than the uncertainties introduced by correcting the Dec-dependence. The conversion to the Baars flux density scale was also to facilitate the use of the GLEAM survey with other radio frequency catalogues and to ensure consistency with future southern hemisphere surveys that will be conducted at frequencies greater than 300\,MHz, such as those to be completed by the Australian Square Kilometre Array Pathfinder \citep[ASKAP;][]{2014PASA...31...41H}.

\subsubsection{Independent test of the GLEAM survey flux density scale}

To independently test the accuracy of the flux density scale of the GLEAM survey, we observed 47~compact sources between Decs of $+25\arcdeg$ and $-45\arcdeg$ using the P-band system on the VLA (proposal 14B-498, PI Callingham). The P-band system is sensitive between 230 and 470\,MHz, providing an overlap at the top of the GLEAM frequency band allowing a direct comparison of the flux density scales. 

\begin{figure*} \begin{center}$
	\begin{array}{cc}
	\includegraphics[scale=0.29]{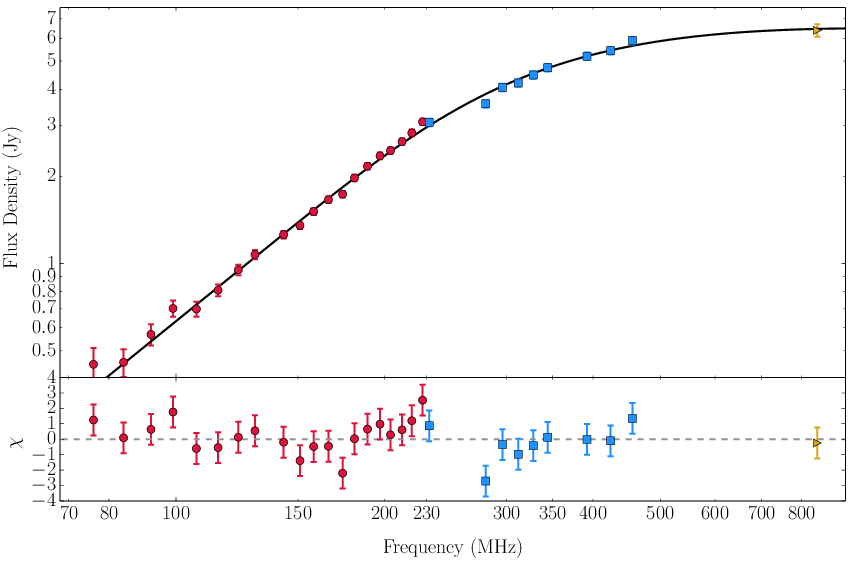} &   
	\includegraphics[scale=0.29]{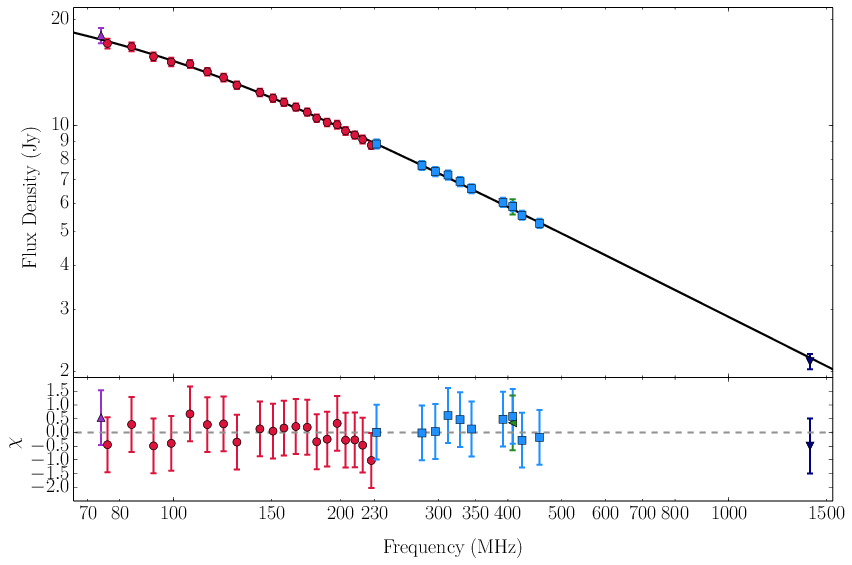}\\
	\end{array}$
	\caption{Spectral energy distributions of two compact sources (left panel: PKS~B0008-421; right panel: PKS~B0310-150) targeted by the VLA to independently test the flux density scale of the GLEAM survey. The P-band and GLEAM data points are shown as blue squares and red circles, respectively. Purple upward-pointing, green leftward-pointing, yellow rightward-pointing, and navy downward-pointing triangles are from the surveys VLSSr, MRC, SUMSS, and NVSS, respectively. Note that a model fit \citep{1997ApJ...485..112B} has been applied to both sources and is shown in black. The $\chi$-values for the model fit to the data are displayed in the panel below the spectral energy distribution.}
    \label{fig:pband_gleam}
  \end{center}
\end{figure*}

The target sources were selected to be compact, non-variable sources that were brighter than 4\,Jy at 408\,MHz, criteria used by the well-characterised MS4 catalogue \citep{2006AJ....131..100B}. Sources were observed in two two-minute snapshot observations with the VLA in either CnB or B configuration. The data reduction was performed using the standard \textsc{AIPS} packages. 3C48 was used to set the flux density scale for the observations, which also placed the P-band observations on the \citet{Baars1977} flux density scale. Any discrepancy between the GLEAM and P-band was always less than a difference of $\sim$5\,\%, well within the uncertainties on the flux density measurements. Therefore, we find no disagreement, within uncertainties, at the top end of the GLEAM band, for the sources in this sample.
The SEDs of two of the sources targeted are provided in \Fig~\ref{fig:pband_gleam}.

\subsubsection{Comparison with other catalogues}\label{sec:TGSS}

Cross-matching the GLEAM extragalactic catalogue with all other overlapping radio surveys is beyond the scope of this paper. However, we make a short comment on one of the most complementary surveys.

The GMRT 150\,MHz All-sky Radio Survey released its First Alternative Data Release \citep[TGSS-ADR1;][]{2016arXiv160304368I} at the time of writing of this paper. As TGSS-ADR1 and GLEAM overlap between $-53\arcdeg<\mathrm{Dec}<+30\arcdeg$, and both make measurements at 150\,MHz, we performed a preliminary comparison of bright ($S_\mathrm{150MHz}>1$\,Jy),
unresolved ($(a\times b)/(a_\mathrm{PSF}\times b_\mathrm{PSF})<1.1$),
and classed as fit by a single Gaussian in TGSS-ADR1) radio sources to compare the two flux density scales. Using these $\approx3,000$ sources, we find the ratio of the GLEAM to TGSS flux density scales is 1.03, averaged over the sky. This 3\% difference is within the range expected since GLEAM and TGSS-ADR1 are on different flux density scales, with TGSS-ARD1 on the scale of \citet{2012MNRAS.423L..30S}, which is a flux density scale bootstrapped from the RCB flux density scale.

In this very early data release, there are significant position-dependent flux density scale discrepancies, which are due to issues with calibration of the GMRT data. At the time of writing, these issues are being solved in order to produce a second data release of TGSS. This first comparison was also used for a transients analysis, detailed in Murphy et al. (submitted). It was also used to remove two false sources from the GLEAM catalogue, which were FFT aliases of bright sources just a few degrees outside the imaged fields-of-view.

Due to the differing resolutions of the two surveys ($25\arcsec$ against $2\arcmin$), fainter sources are more difficult to cross-match correctly. This is even more true for surveys at different frequencies such as SUMSS (843\,MHz; $43\arcsec$), as the apparent morphology of sources can change with frequency. This non-trivial problem will be considered in more detail in the upcoming paper Line et al. (in prep), using suitable Bayesian modelling to discriminate between different potential cross-matches (Line et al. (submitted)).

\subsection{Forming large mosaics}

For each sub-band, each week of observations is combined into a single mosaic, using \textsc{swarp}. This forms four overlapping views of the sky, with some small gaps around very bright sources, and the Galactic Plane. For observations at Dec$=-72\arcdeg$, only pixels which fall within $\pm1$\,hour of RA of the observation RA are used, in order to minimise ionospheric blurring around the South Celestial Pole.
%
%
%
These mosaics form the basis of subsequent analysis. We note that reprojecting all data to a spherical format such as Healpix \citep{2005ApJ...622..759G} allows the combination of all the data in the same image plane. However, we found the week-long mosaics to be the largest useable area without running into computational memory constraints, and that Zenithal Equal Area \citep[ZEA; ][]{2002A+A...395.1077C} is a projection that gives good results in general purpose software, which is especially important for accurate source-finding.

\subsection{Characterising the point spread function}\label{sec:ionopsf}

Typical radio astronomy imaging combines only a few observations in the image plane, for any given pixel. Due to the very wide field-of-view of the MWA, and the drift scan strategy, any given pixel in a mosaic includes contributions from observations over a full hour, during which the ionosphere may distort the positions of sources by varying amounts. The bulk astrometric offsets described in \sect~\ref{sec:astrometry} leave direction-dependent effects of around 5--25\arcsec~, which will be smallest at the centres of each drift scan, where the sensitivity of the primary beam is greatest, and below Dec~$+18\fdg5$, where the MRC catalogue is complete. When we form the week-long mosaics, the lowest frequencies have more overlap than the higher frequencies.

The result of integrating 20--40 observations, with various weights per pixel, tends to be a slight apparent blurring of the point spread function (PSF). Without characterisation and correction, all sources will be detected as slightly resolved by any source-finder, and we will overestimate our ability to resolve real features. This blurring effect varies smoothly over the sky due to the slowly-changing contribution of different observations to each pixel, and increases apparent source areas by 1--25\,\% depending on the frequency of observation and ionospheric activity. The effect on a resulting source catalogue will be to reduce the observed peak flux densities of sources, while integrated flux densities should be preserved. A simple simulation using the typical per-snapshot position difference RMS for each frequency as computed in \Sect~\ref{sec:astrometry} to stack Gaussians in slightly different positions confirms that the blurring we observe is due to the residual ionospheric distortions.

There is also a projection effect, from combining multiple SIN-projected images into a single ZEA mosaic. In a SIN projection, the restoring beam has the correct dimensions across the entire image. When \textsc{swarp} transforms to a ZEA projection, the apparent size of sources is conserved, and so sources with large zenith angle (ZA, $=90\arcdeg - $elevation) appear stretched, taking up more pixels than their zenith equivalents. This is entirely a projection effect which scales as $1/\cos\mathrm{ZA}$. High-ZA sources appear to have larger integrated flux densities while their peak flux densities are preserved. Thus, to correctly characterise the PSF, and correctly flux-calibrate both the peak and integrated flux densities, we must carefully disentangle the projection and blurring effects.

The challenge of measuring the blurring effect is reasonably analogous to optical point-spread-function characterisation, for which several general-purpose packages exist, mainly based on using unresolved sources in the image to sample the shape of the PSF over the image, and performing interpolation over the results. However, unlike in the optical case, in which a population of unresolved stars can be extracted via their sizes and optical colours, it is difficult to distinguish genuinely resolved radio sources from blurred unresolved sources. We use four criteria to distinguish useful unresolved sources which are characteristic of the PSF:
\begin{enumerate}
\item{Not obviously extended: integrated flux density $<2\times$peak flux density;}
\item{Isolated: does not lie within 10\arcmin~ of another source;}
\item{Unresolved in other catalogues: Cross-matches with similarly isolated, point-like MRC ($\mathrm{MFLAG}=0$) or VLSSr ($a$,$b<86$\arcsec) sources;}
\item{Gaussian: residual after subtracting fitted Gaussian is less than 10\,\% of the peak flux.}
\end{enumerate}
Typically this selects 5--15\,\% of the available sources, giving a source density of about one source per square degree.

PSF maps are generated by re-projecting the PSF sample onto a Healpix sphere \citep{2005ApJ...622..759G}, with NSIDE$=16$ (corresponding to $\approx13.4$~square degrees per pixel). Each pixel is averaged together with its neighbours: a form of spherical box-car averaging. During averaging, the sources are weighted by their S/N (peak flux density divided by local RMS measurement) multiplied by their Gaussianity (peak flux density divided by residual of model subtraction). This process forms smoothly-varying maps as a function of position on the sky of major axis, minor axis, and position angle.

In order to restore the peak flux density of the sources, we multiply all of our images by the degree of blurring that we measure, but not the projection effect that arises from the SIN to ZEA transformation. After the PSF map has been measured, its antecedent mosaic is multiplied by a (position-dependent) ``blur'' factor of
\begin{equation}
R = \frac{a_\mathrm{PSF} b_\mathrm{PSF}\cos\mathrm{ZA}}{a_\mathrm{rst} b_\mathrm{rst}}\label{eq:ionoblur}
\end{equation}
where $a_\mathrm{rst}$ and $b_\mathrm{rst}$ are the FWHM of the major and minor axes of the restoring beam. This has the effect of normalising the flux density scale such that both peak and integrated flux densities agree, as long as the correct, position-dependent PSF is used.

\begin{figure*}
	\centering
	\includegraphics[width=\textwidth]{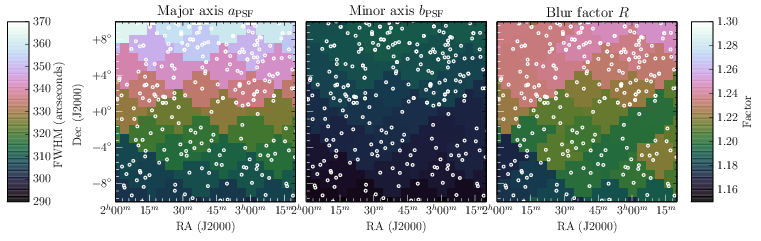} 
	\caption{Point spread function: the major axis (left panel), the minor axis (middle panel), and the blur factor (right panel) for an example $\approx100$~square degrees of the lowest sub-band (72--80\,MHz) image of GLEAM. White ``o''s show the locations of sample sources used to generate the map, at an average source density of one source per square degree. The left colorbar is to be used with the two left panels, and the right colorbar for the right panel only.}
    \label{fig:psfmap}
\end{figure*}

The estimated PSF is stored in the catalogue, and added to the header of any downloaded postage stamp image. This allows users to determine whether a source is really extended, or if there was more ionospheric blurring at that location. It is also used throughout source-finding for the final catalogue: see \Sect~\ref{sec:sourcecat}. An example of the major axis, minor axis, and blur factor for a part of the sky at a single frequency are shown in \fig~\ref{fig:psfmap}. Position angle is not shown as it is entirely determined by projection, and thus is almost always zero (North), except near the zenith where the PSF is circular and thus position angle is meaningless.

\section{Source finding and Cataloguing}
\label{sec:sourcecat}
The cataloguing process is carried out in a tiered approach. For each week of observations a single wideband image is created which covers the frequency range 170--231\,MHz. This image achieves minimal noise and maximum resolution. Sources are extracted from this image and quality control measures are applied to obtain a reference catalogue. The flux density of each source within this reference catalogue is then measured in each of the twenty 7.68\,MHz narrow-band images.

\subsection{Wideband image}
\label{sec:wideband}
The primary product of the image processing described previously is a set of twenty images each with a bandwidth of 7.68\,MHz. In order to construct the most sensitive combined image the following process was used: choose $N$ images starting at the highest frequency (highest resolution) image; convolve all these images to a common resolution (the lowest resolution of the $N$ images); combine these images and measure the noise within this combined image. This process is repeated with a greater $N$ until the noise no longer decreases as more images are added. At this point we have an image with a good compromise between resolution and sensitivity. This process results in a wideband image that covers 170--231\,MHz, with a resolution of $\approx2\arcmin$ (the FWHM of the synthesised beam at 170\,MHz). This wideband image is then used to create a reference catalogue for each of the four observing weeks.

\begin{figure*}
	\centering
	\includegraphics[width=\textwidth]{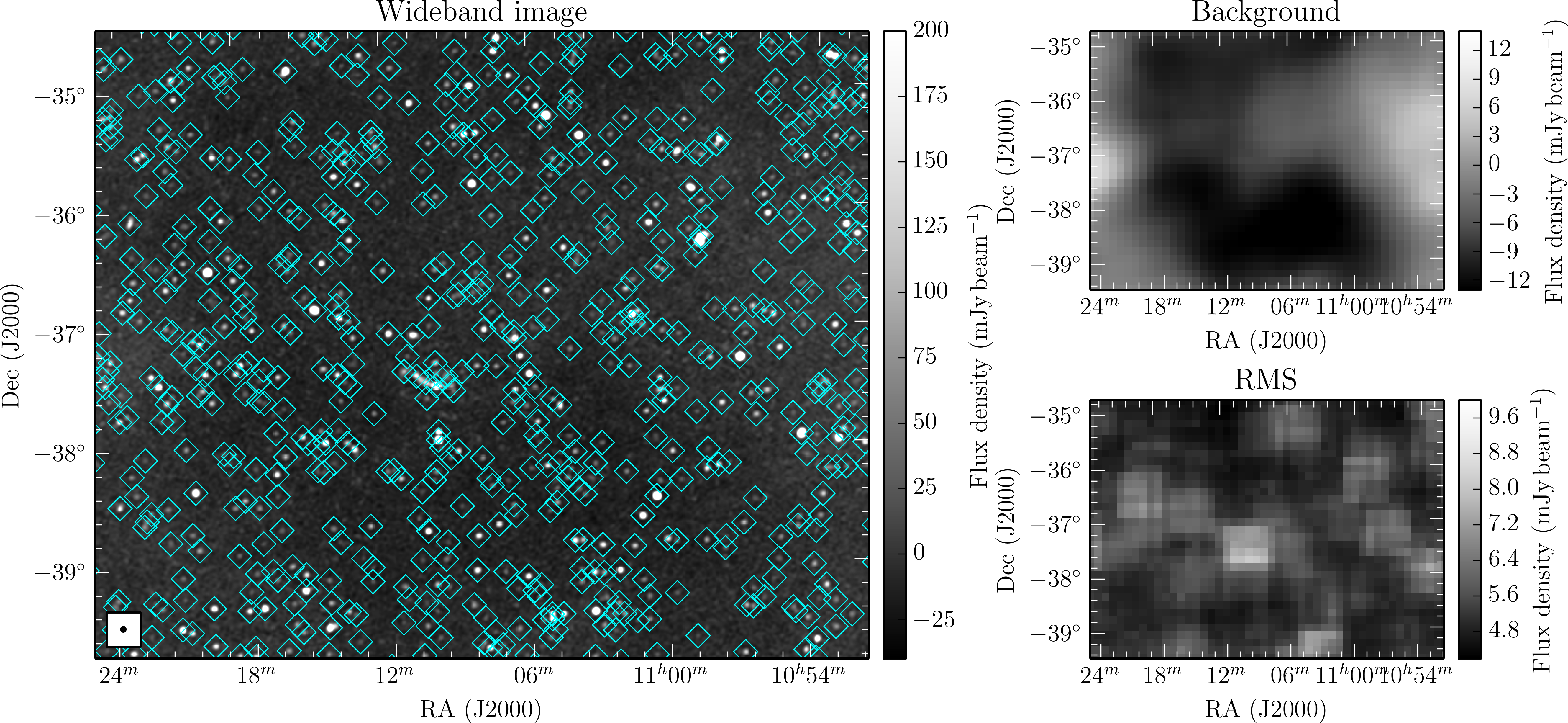}
	\caption{Example 12.5~square degrees from the GLEAM wideband image. The left panel shows the image itself; the top right panel shows the background, and the lower right panel shows the RMS. Detected sources are marked on the main image with cyan diamonds. The PSF for this region is shown as a filled ellipse in the lower left of the main image.}
    \label{fig:example_mosaic}
\end{figure*}

\subsection{Wideband image catalogues}\label{sec:source-finding}
Source finding is performed using \textsc{Aegean}
with a detection threshold of $5\times$ the local RMS. The background emission and RMS noise properties of the wideband images are characterized using \textsc{BANE}, effectively filtering structure of scales $>10\times$ the size of the local PSF into the background image. The point spread function is allowed to vary across the image using the characterization described in the previous section.
A 12.5\,deg$^2$ representative subsection of the wideband image, its measured background, and RMS, are shown in \Fig~\ref{fig:example_mosaic}. Note that because the MWA is sensitive to large-scale Galactic structure, which is not deconvolved, sources appear on a background which can be positive or negative. As the diffuse Galactic synchrotron has a steep spectrum of $\alpha=-2.7$, the background tends to be larger at lower frequencies. This background is subtracted automatically by \textsc{Aegean} during source-finding.

\textsc{Aegean} characterizes sources, or groups of sources, as a combination of elliptical Gaussian components. Each component is described by a position, peak flux density, major and minor axis size and position angle. Each component with the catalogue is assigned a universally unique identifier ({\sc uuid}), which has no meaning in and of itself, but plays an important role in the matching of sources in narrow-band images.

\begin{figure*}
	\centering
	\includegraphics[width=0.8\textwidth]{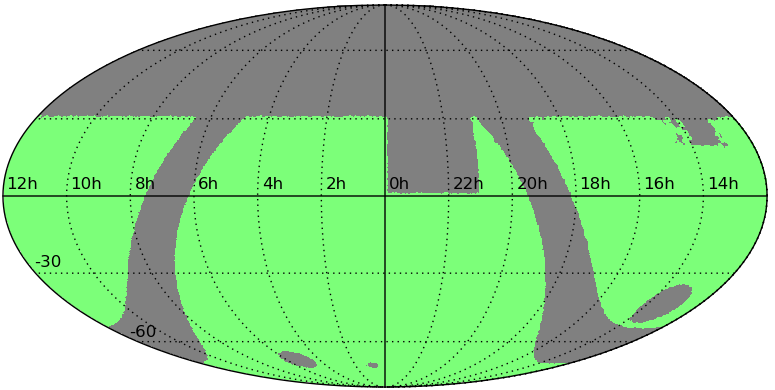}
	\caption{Detected sources that fall within the green shaded area, described in \Tab~\ref{tab:excluded_regions}, are included in the catalogue.}
    \label{fig:footprint}
\end{figure*}

A number of position-based filters were implemented in order to remove false detections and sky areas that are beyond the scope of this paper. Sources that fell: within $10\arcdeg$ of the Galactic plane, within $5\fdg5$ and $2\fdg5$ of the Large or Small Magellanic Clouds, respectively, within 10\arcmin~ of peeled sources (\Tab~\ref{tab:calibrators}), within 9\arcmin~ of Centaurus~A, or North of Dec$+30\arcdeg$, were removed from the catalogues. A northern region was also discarded due to two nights of high ionospheric activity in the first week of observing, and a further northern region was discarded because Centaurus~A fell in a primary beam side-lobe and made the self-calibration stage as designed impossible. \Fig~\ref{fig:footprint} shows the footprint of the survey region after these positional filters were applied, and \Tab~\ref{tab:excluded_regions} lists the exclusion zones.

Note that ionospheric blurring affects the resolution of the wideband image the least, of order 5\,\%, due to the high frequency of its component observations. Thus, source morphologies are well-characterised, despite the residual effects of ionospheric distortion.

\begin{table*}
	\caption{The areas surveyed (top row), flagged (middle rows), and catalogued (final row) in this paper. $r$ indicates the radius around a source inside which flagging was performed.}
	\begin{tabular}{ccc}
	\hline
	Description & Region & Area / square degrees\\
	\hline
        \textbf{Total surveyed area} & $\mathrm{Dec}<+30\arcdeg$ & \textbf{30,940} \\
	\hline
	Galactic plane & Absolute Galactic latitude $<10\arcdeg$ & 4,776 \\
    Ionospherically distorted & $0\arcdeg<\mathrm{Dec}<+30\arcdeg$ \& $22^\mathrm{h}<\mathrm{RA}<0^\mathrm{h}$ & 859 \\
	Centaurus~A (see \Tab~\ref{tab:calibrators}) & $13^\mathrm{h}25^\mathrm{m}28^\mathrm{s} -43\arcdeg01\arcmin09\arcsec$, $r=9\arcdeg$ & 254 \\
    Sidelobe reflection of Cen~A & $13^\mathrm{h}07^\mathrm{m}<\mathrm{RA}<13^\mathrm{h}53^\mathrm{m}$ \& $20\arcdeg<\mathrm{Dec}<+30\arcdeg$ & 104 \\
	Large Magellanic Cloud & $05^\mathrm{h}23^\mathrm{m}35^\mathrm{s} -69\arcdeg45\arcmin22\arcsec$, $r=5\fdg5$ & 95 \\
	Small Magellanic Cloud & $00^\mathrm{h}52^\mathrm{m}38^\mathrm{s} -72\arcdeg48\arcmin01\arcsec$, $r=2\fdg5$ & 20 \\
	Peeled sources (see \Tab~\ref{tab:calibrators}) & $r=10\arcmin$ & $<1$  \\
	\hline
        \textbf{Final catalogue area} & & \textbf{\survarea} \\
	\hline
	\end{tabular}
	\label{tab:excluded_regions}
\end{table*}

\subsection{Narrow-band image catalogues}\label{sec:priorised}
The catalogue entries for the narrow-band images are not created via blind source finding. For each source in the reference catalogue we measure the flux density in each of the narrow-band images. We call this measurement process {\em priorised fitting}; we utilise \textsc{Aegean}, and outline the processing here.

Each of the narrow-band images has a different resolution and so the measurement process begins by determining the expected shape of the sources from the reference catalogue. A source from the reference catalogue is deconvolved by the local PSF in the wide-band image, and then convolved with the local PSF from the narrow-band image. All sources within the reference catalogue are then sorted into groups such that any sources that overlap at the half-power point of their respective Gaussian fits are put into the same group. A fit is then performed for the peak flux of each source, with the position and newly-determined shape parameters held fixed. This fit is performed over all sources within a group at the same time. Rarely, it is not possible to make a measurement of a source in a narrow-band image. This can occur if the local PSF was not able to be determined for that image, or if part of the sky was not able to be imaged at a particular frequency. Just under 2\,\% of sources do not have a measurement in one or more sub-bands.

The fitted and fixed source parameters are recorded and each source is assigned the same {\sc uuid} as its corresponding reference source. The process of associating sources from the narrow-band images with their reference sources within the wideband images is achieved by matching {\sc uuid}s. This refitting and matching process guarantees the extraction of intra-band spectral energy distributions for all sources, without having to rely on position-based cross-matching of catalogues that may describe a single source with a different number of components in each of the narrow-band images. Since the narrow-band image measurement process does not involve blind source finding, there is no signal-to-noise cut placed on the fluxes from these narrow-band images. As a result, it is possible for the reported flux in the narrow-band images to be less than the RMS noise level, or even negative. The presence or absence of a flux density measurement at a narrow-band frequency does not indicate a detection or non-detection, but merely that a measurement was made. We note that this also avoids overestimation of the flux density of faint sources in the narrow-band images, because their shape parameters are not allowed to vary, so cannot be extended by a local increase in the RMS noise.

\subsection{Final catalogue}
Once the narrow-band catalogues have been created and curated, they are combined together to generate the GLEAM master catalogue. Where there was overlap between different observing weeks, the sky area with the lower RMS noise was chosen to produce the master catalogue. This catalogue lists the location of each source as measured in the wide-band image, the integrated flux, and shape of each source at each of the frequencies within the survey, along with the local PSF at the location of each source at each frequency.
The catalogue contains $\nsrc$~rows, and $\ncol$ columns. Column names, units, and descriptions are shown in \Tab~\ref{tab:appendix}. The electronic version of the full catalogue is available from VizieR.

\subsection{Error derivation}

In this section we examine the errors reported in the GLEAM catalogue. First, we examine the systematic flux density errors from our primary beam mirroring technique. Then, we examine the noise properties of the wide-band source-finding image, as this must be close to Gaussian in order for sources to be accurately characterised, and for estimates of the reliability to be made, which we do in \Sect~\ref{sec:reliability}. Finally, we make an assessment of the catalogue's astrometric accuracy. These statistics are given in \Tab~\ref{tab:survey_stats}.

\subsubsection{Flux density scale uncertainty}\label{sec:systematic-error}

For the majority of the GLEAM survey, the dominating uncertainty in the flux density measurements results from the Dec-dependent flux density correction. This systematic uncertainty is due to the spread in correction factors, as is evident in \Fig~\ref{fig:flux_scale_dec-13} and discussed in \Sect\,\ref{sec:corr_pb}. The systematic uncertainty is calculated as the standard deviation of a Gaussian fit to the remaining variation in the ratio of predicted to measured source flux densities, as shown in the right panel of ~\Fig~\ref{fig:correction_survey}.

For sources lying between Decs $-72\arcdeg$ and $+18\fdg5$, these uncertainties are $8\pm0.5$\,\% of the integrated flux density. As MRC only has coverage up to a Dec of $+18\fdg5$, the systematic uncertainties are larger, $11\pm2$\,\%, for sources with a Dec greater than $+18.5\arcdeg$ and those with a Dec between $-83\fdg5$ and $-72\arcdeg$, the latter due to the primary beam mirroring technique (\Sect~\ref{sec:corr_pb}).
Finally, sources with $\mathrm{Dec}<-83\fdg5$, near the South Celestial Pole, have close to $\sim$80\,\% systematic uncertainties as the flux density polynomial corrections used for this area are mirrored extrapolations from Dec $30\arcdeg$--$36\fdg6$ (\Fig~\ref{fig:fan_diagram}).
The catalogue therefore contains a column indicating the expected systematic uncertainty for each source, based on its Dec: 80\,\% for $-90\arcdeg\leq\mathrm{Dec}<-83\fdg5$ (911~sources); 13\,\% for $-83\fdg5\leq\mathrm{Dec}<-72\arcdeg$ (8,821~sources) and $+18\fdg5\leq\mathrm{Dec}<30\arcdeg$ (15,452~sources); and 8\,\% for $-72\arcdeg\leq\mathrm{Dec}<18\fdg5$ (280,431~sources).

\subsubsection{Noise properties}\label{sec:noise_properties}
Here we examine the characteristics of the noise in the wideband image. We use a 675\,deg$^2$ region where there are a fairly typical number of bright sources (20 with $S>5$\,Jy) and the mean RMS noise is quite low ($7.6$\,mJy\,beam$^{-1}$), centred on RA 3$^\mathrm{h}$, Dec $-27\arcdeg$.

\begin{figure*}
	\centering
	\includegraphics[width=0.9\textwidth]{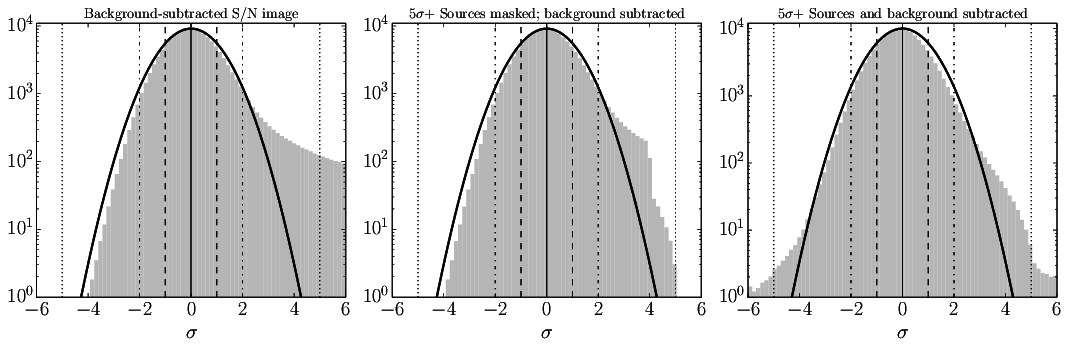} 
	\caption{Noise distribution in a typical 675\,square degrees of the wideband source-finding image. BANE measures the average RMS in this region to be 7.6\,mJy\,beam$^{-1}$. To show the deviation from Gaussianity, the ordinate is plotted on a log scale. The leftmost panel shows the distribution of the S/Ns of the pixels in the image produced by subtracting the background and dividing by the RMS map measured by \textsc{BANE}; the middle panel shows the S/N distribution after masking all ($S>4\sigma$) pixels over which sources were characterised; the right panel shows the S/N distribution after subtracting the models of the detected sources. The light grey histograms show the data. The black lines show Gaussians with $\sigma=1$; vertical solid lines indicate the mean values. $|\mathrm{S/N}|=1\sigma$ is shown with dashed lines, $|\mathrm{S/N}|=2\sigma$ is shown with dash-dotted lines, and $|\mathrm{S/N}|=5\sigma$ is shown with dotted lines. The paucity of negative pixels underneath the Gaussians shows that BANE over-estimates the noise in this region by $\approx7.5$\,\%, due to the strong confusion component (grey pixels on the right-hand side of the distribution in the middle pannel). }
    \label{fig:gaussian_noise}
\end{figure*}

To characterise the area outside of detected sources, we use two different methods: masking the ($S>4\sigma$) pixels which are used during source characterisation; and subtracting the measured source models from the image. Since the RMS noise varies over the image, we divide the resulting images by the original noise images to produce images of S/N, and plot the distributions of these pixels in \Fig~\ref{fig:gaussian_noise}. The original image has a negative mean due to the undeconvolved sidelobes of the diffuse Galactic background; this has been subtracted from the plotted distributions.

For the masked image, no pixels reach $|\mathrm{S/N}|>5\sigma$, which is consistent with the source-detection algorithm.
There is also a surfeit of pixels with $\mathrm{S/N}>2.5\sigma$, which is unsurprising, because these pixels include the fainter tails of sources not included when the sources are characterised, as well as many faint, real, confused sources. 
The distribution is thus slightly asymmetric; \textsc{BANE} has attempted to determine a characteristic RMS noise, but when considering the negative pixels, this RMS noise appears to be an overestimate of approximately 7.5\,\%. This is due to its use of sigma-clipping, which does not cope well with the large number of 3--$5\sigma$ confused sources present at this noise level. This is difficult to correct for, as the noise level varies over the sky, and devising a new noise estimator is beyond the scope of this paper. Therefore we note that the images may contain believable sources which will not appear in the catalogue; this is preferential to the reverse situtation of underestimating the noise, and lowering the reliability of the catalogue.

For the residual image, there is both a negative and positive tail. Note that these pixels are within the extents of detected sources. They are caused by imperfect modelling of sources using elliptical Gaussians, which is not unexpected given that sources may have real extent, and calibration errors act to make sources less Gaussian. We expect any unreal sources to lie only within $6\arcmin~$ of detected sources, although calibration errors around extremely bright ($S>100$\,Jy) sources are not considered in this particular region of sky (see \Sect~\ref{sec:reliability} for a reliability analysis of the whole sky).

%
We note here that it is the sidelobe confusion in particular which limits the depth of this survey; after an effective integration time of $\approx10$\,minutes, some areas of the wideband image (particularly near zenith, where the primary beam has most sensitivity) reach RMS noise levels of 5\,mJy\,beam$^{-1}$. However, further integration time would not significantly reduce the noise, if we continue to individually deconvolve each snapshot to a $3\sigma$ threshold, because fainter sources will never be \textsc{clean}ed. Instead, it is necessary to use a peeling strategy such as that adopted by \cite{2016MNRAS.458.1057O}. Eventually one will reach the classical confusion limit, measured by \cite{2016MNRAS.459.3314F} to be $\simeq1$\,mJy\,beam$^{-1}$ for the MWA, at these frequencies. \cite{2015mfer.confE..53F} describes the confusion properties of the GLEAM survey across the full 72--231\,MHz bandwidth: the confusion at 231\,MHz is $<1$\,mJy, $<3$\,\% of the typical local RMS at that frequency, while the confusion at 72\,MHz is $\approx10$\,mJy, $\approx10$\,\% of the typical local RMS at that frequency.

\subsubsection{Completeness}

The completeness of GLEAM cannot be quantified using existing surveys, due to its unique combination of high surface brightness sensitivity, low frequency of measurement, and relatively low resolution. As detailed in \Sect~\ref{sec:TGSS}, GLEAM was cross-matched with TGSS-ADR1 to verify its flux density scale; unfortunately, this early data release suffered from position-dependent calibration errors, which mean its completeness is quite unknown, so it cannot serve as a reference for the completeness of GLEAM. NVSS and SUMSS are more sensitive surveys, but for a physically-reasonable spectral index limit of $\alpha=-2.5$, their best completeness limits of $S_\mathrm{1.4GHz}=2.5$\,mJy and $S_\mathrm{843MHz}=8$\,mJy are, at 200\,MHz, only 324\,mJy and 292\,mJy, respectively. As nearly 80\,\% of the GLEAM catalogue sources have flux densities $<300$\,mJy, and an unknown number of these sources could have spectral indices $\simeq-2.5$, these higher-frequency surveys cannot be used as a reference for the completeness of GLEAM. There is also the difficult issue of cross-matching surveys of very different resolutions; this problem will be addressed in the upcoming paper Line et al. (in prep).

Instead, simulations are used to quantify the completeness of the GLEAM source catalogue at 200\,MHz.
It is beyond our computational budget to perform these simulations on the individual observations, so the characterisation is performed after flux-calibrated mosaics have been formed.
Thirty-three realisations were used in which 250,000 simulated point sources of the same flux density were injected into the 170--231\,MHz week-long mosaics. 
The flux density of the simulated sources is different for each realisation, spanning the range 25~mJy to 1~Jy.
The positions of the simulated sources are chosen randomly but not altered between realisations; to avoid introducing an artificial factor of confusion in the simulations, simulated sources
were not permitted to lie within 10\arcmin~ of each other.

Sources are injected into the week-long mosaics using \textsc{aeres} from the \textsc{aegean} package.
Areas flagged from the GLEAM source catalogue (see Table~\ref{tab:excluded_regions}) are excluded from the simulations.
The major and minor axes of the simulated sources are set to $a_\mathrm{psf}$ and $b_\mathrm{psf}$, respectively.

For each realisation, the source-finding procedures described in \Sect~\ref{sec:source-finding} are applied to the mosaics and the fraction of simulated sources recovered is calculated.
In cases where a simulated source is found to lie too close to a real ($>5\sigma$) source to be detected separately, the simulated source is considered to be detected if the recovered source position is closer to the simulated rather than the real source position. This type of completeness simulation therefore accounts for sources that are omitted from the source-finding process through being too close to a brighter source.

\begin{figure}
\begin{center}
\includegraphics[width=0.5\textwidth]{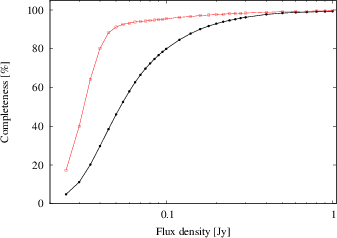} 
\caption {Estimated completeness of the GLEAM source catalogue as a function of $S_{200 \mathrm{MHz}}$
in the entire survey area (black circles) and in the region used to measure the source counts (red squares; see \Sect~\ref{sec:source_counts}).}
\label{fig:cmp_vs_s}
\end{center}
\end{figure}

\fig~\ref{fig:cmp_vs_s} shows the fraction of simulated sources recovered as a function of 
$S_{200 \mathrm{MHz}}$ in the entire survey area.
The completeness is estimated to be 50\,\% at $\approx 55$\,mJy rising to
90\,\% at $\approx 170$\,mJy. \fig~\ref{fig:cmp_vs_s} also shows the completeness as a function of $S_{200 \mathrm{MHz}}$ 
in the most sensitive areas of the survey ($0^\mathrm{h}<\mathrm{RA}<3^\mathrm{h}$ and $-60\arcdeg<\mathrm{Dec}<-10\arcdeg$; $10^\mathrm{h}<\mathrm{RA}<12^\mathrm{h}$ and $-40\arcdeg<\mathrm{Dec}<-15\arcdeg$) where the noise is uniform ($6.8\pm1.3$\,mJy\,beam$^{-1}$).
The completeness in these areas is estimated to be 50\,\% at $\approx 34$\,mJy and 90\,\% at $\approx 55$\,mJy. Errors on the completeness estimate are derived assuming Poisson errors on the number of simulated sources detected.

\begin{table*}
	\caption{Survey properties and statistics. We divide the survey into four Dec ranges, as shown in \Fig~\ref{fig:fan_diagram}, because the noise properties, and astrometric and flux calibration, differ slightly for each range. Values are given as the mean$\pm$the standard deviation. The statistics shown are derived from the wideband (200\,MHz) image. The flux scale error applies to all frequencies, and shows the degree to which GLEAM agrees with other published surveys. The internal flux scale error also applies to all frequencies, and shows the internal consistency of the flux scale within GLEAM.\label{tab:survey_stats}}
	\begin{tabular}{ccccc}
	\hline
	Property & $\mathrm{Dec}<-83\fdg5$ & $-83\fdg5\leq\mathrm{Dec}<-72\arcdeg$ & $-72\arcdeg\leq\mathrm{Dec}<+18\fdg5$ & $\mathrm{Dec}\geq18\fdg5$ \\
	\hline
    Number of sources & 920 & 8,780 & 281,931 & 16,170 \\
    RA astrometric offset (\arcsec) & $-4\pm16$ & $-4\pm16$ & $-0.2\pm3.3$ & $0.5\pm2.5$ \\
    Dec astrometric offset (\arcsec) & $0.1\pm3.6$ & $-0.1\pm3.6$ & $-1.6\pm3.3$ & $1.7\pm2.7$ \\
    External flux scale error (\%) & 80 & 13 & 8 & 13 \\
    Internal flux scale error (\%) & 3 & 3 & 2 & 3 \\
    Completeness at 50\,mJy (\%) & 10 & 22 & 54 & 3 \\
    Completeness at 100\,mJy (\%) & 81 & 83 & 87 & 30 \\
    Completeness at 160\,mJy (\%) & 96 & 95 & 95 & 56 \\
    Completeness at 0.5\,Jy (\%) & 99 & 99 & 99 & 94 \\
    Completeness at 1\,Jy (\%) & 100 & 100 & 100 & 97 \\
    RMS (mJy\,beam$^{-1}$) & $23\pm7$ & $15\pm5$ & $10\pm5$ & $28\pm18$ \\
    PSF major axis (\arcsec) & $196\pm8$ & $176\pm8$ & $140\pm10$ &  $192\pm14$ \\
    PSF minor axis (\arcsec) & $157\pm9$ & $149\pm8$ & $131\pm4$ & $135\pm2$ \\
	\hline
    \end{tabular}
\end{table*}

%

The survey completeness varies substantially across the sky because of the presence of bright sources and varying observational data quality.
In order to map the variation of the completeness across the sky, we have produced maps of the completeness at flux density levels from 25 to 1000\,mJy.
The completeness at any pixel position is given by $C = N_{\mathrm{d}}/N_{\mathrm{s}}$, where $N_{\mathrm{s}}$ is the number of simulated sources in a circle of radius $6\arcdeg$ centred on the pixel and $N_{\mathrm{d}}$ is the number of simulated sources that were detected above $5\sigma$ within this same region of sky.
The completeness maps, in \textsc{fits} format, can be obtained from the supplementary material. Postage stamp images from our VO server also include this completeness information in their headers.

Assuming Poisson statistics, the error on the completeness, $\delta C$, is given by $\sqrt{N_{\mathrm{d}}}/N_{\mathrm{s}} = \sqrt{C/N_{\mathrm{s}}}$.
Given that 200,020 sources were randomly distributed over an area of \survarea\,$\mathrm{deg}^{2}$,
$N_{\mathrm{s}} \approx 900$.
Therefore, $\delta C \approx \sqrt{C}/30$. For example, for $C = 0.9$, $\delta C \approx 0.03$ and for $C = 0.5$, $\delta C \approx 0.02$.

\begin{figure}
\begin{center}
\includegraphics[width=0.5\textwidth]{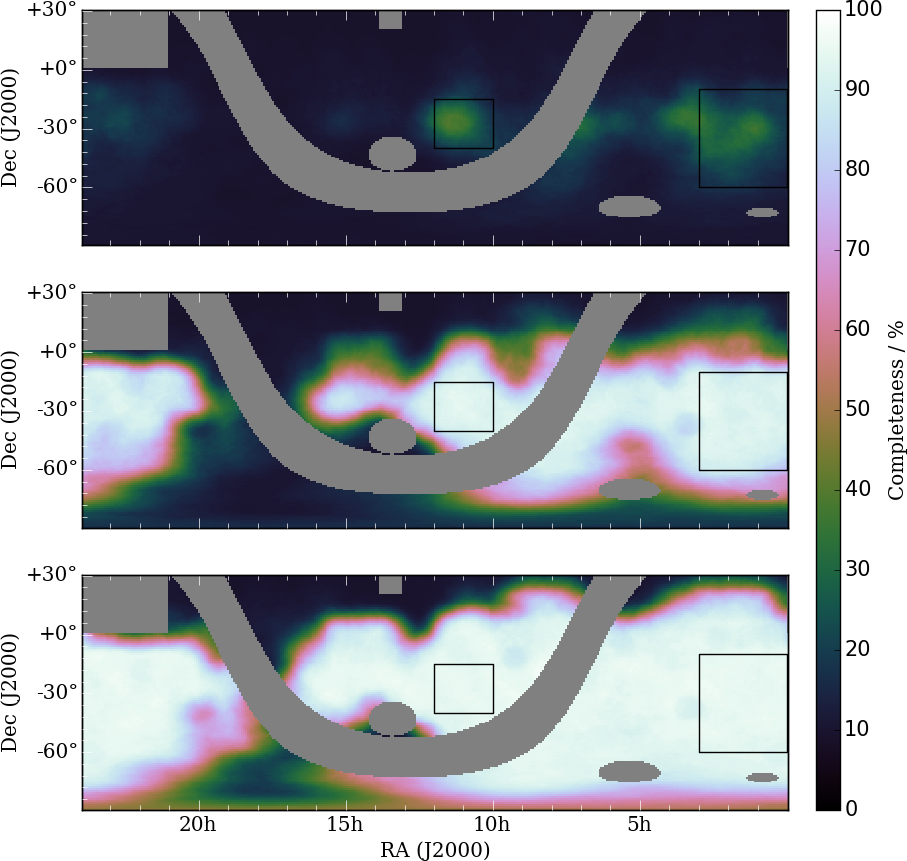}
\caption{The top, middle and bottom panels show completeness maps at 25, 50 and 75\,mJy, respectively, for the regions covered
by the GLEAM catalogue of this paper. Areas which are excluded from the survey are shaded grey.
The image projection used is Cartesian. The most sensitive areas of the survey in which the completeness was calculated
in \Fig~\ref{fig:cmp_vs_s} are outlined in black. \textsc{FITS} images of these completeness maps are available online.
}
\label{fig:cmp_map}
\end{center}
\end{figure}

\Fig~\ref{fig:cmp_map} shows completeness maps at 25, 50 and 75\,mJy.
The completeness is highest close to the zenith.
At 50\,mJy, the completeness is $\approx 90$\,\% at most RAs. There are regions where the completeness at 75\,mJy remains poor, 
particularly at high Decs, due to the presence of bright contaminating sources, rapid sky rotation, and reduced primary beam sensitivity.

\subsubsection{Reliability}\label{sec:reliability}

In order to estimate the reliability of the GLEAM catalogue, we run the source finder \textsc{Aegean} on the four week-long mosaics covering 170--231~MHz in exactly the same way as described in \Sect~\ref{sec:sourcecat}, except that we used a special mode of \textsc{Aegean}
that only reports sources with negative flux densities.
After applying position-based filtering (\Tab~\ref{tab:excluded_regions}), the 
total number of sources with negative peaks below $-5 \sigma$, hereafter referred to as ``negative''; sources, was 97.
The most negative recorded peak flux density was $-8.3 \sigma$. 
If the noise distribution were symmetric about zero, we would expect a number of spurious sources identical to the number of ``negative'' sources.
The total number of sources above $5 \sigma$ in the GLEAM catalogue is \nsrc.
Using the argument that there should be as many false positives as detected negative sources, we estimate the reliability to be $1.0 - \frac{97}{\nsrc} = {\pctreliable}$\,\%. We note that since priorised fitting was 
used to characterise sources in the subband images, the reliability is the same in all subband images.

The majority of the negative sources lie close to bright sources and result from image artefacts
caused by calibration and deconvolution errors: 67 (69\,\%) of the negative sources lie
within 6\arcmin~ of a source brighter than 3\,Jy. Sources with S/Ns $\lesssim 10$ lying within a few arcmin from strong ($\gtrsim 3$~Jy) sources are most likely to be spurious.

\subsubsection{Astrometry}\label{sec:overall_astrometry}

We measure the astrometry using the 200-MHz catalogue, as this provides the locations and morphologies of all sources in the catalogue. To determine the astrometry,
unresolved ($(a\times b)/(a_\mathrm{PSF}\times b_\mathrm{PSF})<1.1$),
isolated (no internal match within 10\arcmin) GLEAM sources are cross-matched with similarly isolated NVSS and the Sydney University Molonglo Sky Survey \citep[SUMSS;][]{1999AJ....117.1578B}; the positions of sources in these catalogues are assumed to be correct and RA and Dec offsets are measured with respect to those positions. In the well-calibrated Dec range of $-72\arcdeg\leqslant\mathrm{Dec}\leqslant+18\fdg5$, the average RA offset is $-0\farcs2\pm3\farcs3$, and the average Dec offset is $-1\farcs6\pm3\farcs3$. North of $+18\fdg5$, the average RA offset is $0\farcs5\pm2\farcs5$ and the average Dec offset is $1\farcs7\pm2\farcs7$. These offsets may be somewhat different because a modified VLSSr/NVSS catalogue was used to replace MRC North of its Declination limit of 18\fdg5 (see \sect~\ref{sec:astrometry}).

Moving south of Dec$-72\arcdeg$, the average RA offset deteriorates to $-4\arcsec\pm16\arcsec$, while the Dec offset remains reasonable at $-0\farcs1\pm3\farcs6$. The RA is particularly distorted because the data have been averaged in the image plane over many hours, on the edge of the field-of-view, where the ionospheric corrections are poorest. This preferentially smears out the sources in hour angle, or RA, direction.

\begin{figure}
	\centering
	\includegraphics[width=0.5\textwidth]{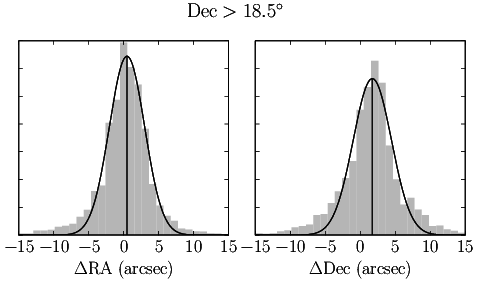} 
	\includegraphics[width=0.5\textwidth]{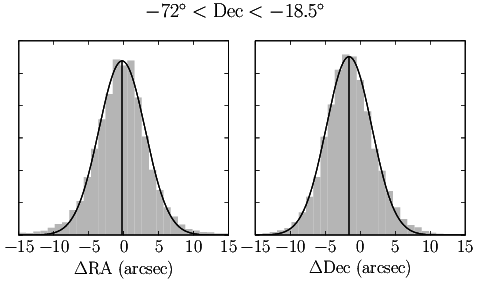}
	\includegraphics[width=0.5\textwidth]{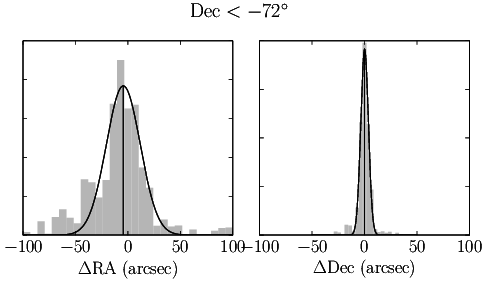}
	\caption{Histograms, weighted by source S/N, of astrometric offsets, for isolated compact GLEAM sources crossmatched with NVSS and SUMSS as described in \sect~\ref{sec:overall_astrometry}. The black curves show Gaussian fits to each histogram. Solid vertical lines indicate the mean offsets. The top panel shows sources on the northern edge of the survey, $\mathrm{Dec}\geq+18\fdg5$; the middle panel shows sources in the main area of the survey, $-72\leq\mathrm{Dec}<+18\fdg5$; and the lowest panel shows sources near the South Celestial Pole, $\mathrm{Dec}<-72\arcdeg$. Note that the range on the abscissa changes for the lowest panel.}
    \label{fig:overall_astrometry}
\end{figure}

For 99\% of sources, fitting errors are larger than the measured average astrometric offsets. Given the scatter in the measurements, and the small relative size of the worst-affected regions, we do not attempt to make a correction for these offsets. Given that we have corrected each snapshot, residual errors should not vary on scales smaller than the size of the primary beam. We advise users to use particular caution when crossmatching sources south of Dec~$-72$\arcdeg with other catalogues. \fig~\ref{fig:overall_astrometry} shows the density distribution of the astrometric offsets, and Gaussian fits to the RA and Dec offsets.

\subsection{Resolved sources}
Only objects that can be described well by one or more elliptical Gaussians have been included in this catalogue; as described in \sect~\ref{sec:sourcecat}, highly-resolved, diffuse sources are excluded. We fit \nresolved ``resolved'' sources, as defined by having a size ($a\times b$) more than 10\,\% greater than the local PSF. Our fitting procedure deals with multi-component sources by fitting the components simultaneously, if their proximity warrants it. As a guide, \nclustered sources lie within 6\arcmin~ of another source. Diffuse, extended, or steep-spectrum structure visible only at the lowest frequencies will be missing from the catalogue, as all source selection is performed at 200\,MHz, and characterisation is performed by extrapolating the shape of the source using the difference in frequency and the measured PSF.

\section{Properties of the catalogue}\label{sec:properties}
Here we compare the catalogue to other MWA data products, measure the spectral indices of its sources, analyse its source counts, and discuss the challenges associated with spectral model fitting with GLEAM data. The upcoming paper White et al. (in prep) calculates a two-point angular correlation function for the catalogue.

\subsection{GLEAM, MWACS and the EoR field}

We crossmatch GLEAM with the MWA Commissioning Survey at 180\,MHz \citep[MWACS; ][]{2014PASA...31...45H}
and the deep 163\,MHz catalogue of the ``zeroth'' Epoch of Reionisation (EoR) field centred at RA$=00^\mathrm{h}$, Dec$=-26\fdg7$ by \cite{2016MNRAS.458.1057O}, the flux scale of which is largely set by MWACS. For sources with $S>1$\,Jy at 180\,MHz, we find that GLEAM has a flux scale 15\,\%($\pm8\%$) lower than both of these catalogues. This is likely due to residual primary beam errors in MWACS, which did not use as accurate a beam model, or fit a model to the residual flux errors as in \Sect~\ref{sec:corr_pb}. We note that the flux scale scatter shown in \Fig~\ref{fig:correction_survey} is also a 2\% improvement on the flux scale scatter of MWACS, shown in \Fig~11 of \cite{2014PASA...31...45H}.

\subsection{Spectral index distribution}\label{sec:alpha}

The wide bandwidth of GLEAM allows an internal spectral index calculation; 98\,\% of sources have a flux density measurement in every sub-band; 75\,\% of sources have a non-negative flux density measurement in every sub-band. For these \npos sources, we calculate $\alpha$ using a weighted least-squares approach.

While a flux density scale error of 8\,\% is needed to reconcile GLEAM with other surveys (\sect~\ref{sec:systematic-error}), an error of 2\,\% gives more consistent results when fitting spectral indices using only the GLEAM data. Specifically, the median value of the reduced $\chi^2$ statistic, which should be unity for data with well-estimated error bars, is biased low at high flux densities for a flux density scale error of 8\,\%, while it is unity across the catalogue for a flux density scale error of 2\,\%. We note that estimating the error in this way is incorrect for a \textit{single} measurement, but is not problematic for such a large catalogue. This error increases to 3\,\% for $\mathrm{Dec}>18\fdg5$ and $\mathrm{Dec}<-72\arcdeg$.

That the internal flux scale consistency is better than the external is likely due to a combination of factors: the narrow time-frame in which the observations were taken, minimising astrophysical variability; the consistency of using a single instrument with near-uniform spatial frequency sensitivity and calibration rather than a variety of instruments; the wide field-of-view and large amount of snapshot averaging acting to suppress any flux scale variations on small scales; and, most importantly, the scale being derived from identical sources, identical surveys, and from similar large-angular-scale polynomials (\Sect~\ref{sec:fluxscale}). Therefore, during the spectral energy distribution fitting process, we set the error on each flux density to be the quadrature sum of the \textsc{Aegean} fitting error, and a 2\,\% (3,\%, for $\mathrm{Dec}>18\fdg5$ and $\mathrm{Dec}<-72\arcdeg$) internal flux scale error.

Note that fitting a single power law spectrum ignores potentially interesting astrophysics, such as jet confinement or synchrotron self-absorption giving rise to sources with peaked or flattening spectra. The upcoming paper Callingham et al. (in prep) contains more extensive spectral modelling of such sources found in this catalogue. To exclude poorly-fit spectral indices from the catalogue presented in this work, and the spectral indices presented in this section, we require that $\chi^2<34.805$, as $P(\chi^2 \geq 34.805)>99\%$ for 18~degrees of freedom (i.e. reduced~$\chi^2<1.93$). This results in calculated spectral indices for \nfit sources.

\begin{figure}
	\centering
	\includegraphics[width=0.5\textwidth]{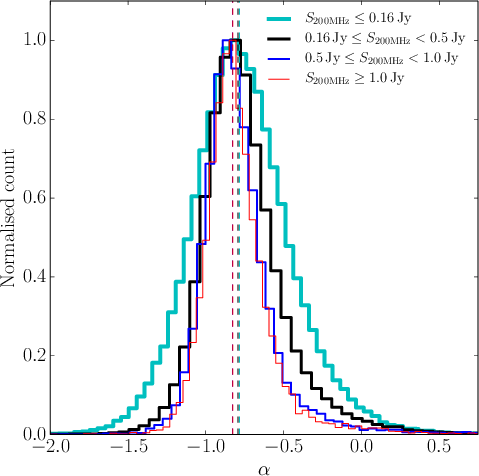} 
	\caption{The spectral index distribution calculated using solely the catalogue described in this paper. The
cyan line shows sources with $S_\mathrm{200MHz}<0.16$\,Jy, the
black line shows sources with $0.16\leq S_\mathrm{200MHz}<0.5$\,Jy, the blue line shows sources with $0.5\leq S_\mathrm{200MHz}<1.0$\,Jy, and the red line shows sources with $S_\mathrm{200MHz}>1.0$\,Jy. The dashed vertical lines of the same colours show the median values for each flux density cut: $-0.78$, $-0.79$, $-0.83$, and $-0.83$, respectively.
    \label{fig:alpha_distribution}}
\end{figure}

The resulting distributions of $\alpha$ for sources with $S_\mathrm{200MHz}<0.16$\,Jy (122,959~sources), $0.16\leq S_\mathrm{200MHz}<0.5$\,Jy (86,548~sources), $0.5\leq S_\mathrm{200MHz}<1.0$\,Jy (20,606~sources), and $S_\mathrm{200MHz}\geq1.0$\,Jy (12,723~sources), are shown in \Fig~\ref{fig:alpha_distribution}. The median and semi-inter-quartile-range (SIQR) values of $\alpha$ for these flux density bins are $-0.78\pm0.20$, $-0.79\pm0.15$, $-0.83\pm0.12$, and $-0.83\pm0.11$, respectively.

Note that only the brightest two bins are substantially complete over the whole sky, as shown in Table~\ref{tab:survey_stats}. The errors on the flux densities in the faintest bin are of order 30\,\%, and all sources with negative flux density measurements in sub-bands are discarded, so the wide distribution and slightly flatter average spectral index of the faintest flux density bin should not be interpreted as a properties intrinsic to those sources.

These values are in very good agreement with other measurements at these frequencies; for instance, VLSSr and NVSS together measure $\alpha=-0.82$ with an SIQR of 0.11 \citep{2014MNRAS.440..327L}, while in the MSSS verification field, \cite{2015A&A...582A.123H} use the LOFAR High Band Antenna (120--160\,MHz) to measure a median $\alpha$ of $-0.77$. \cite{2003MNRAS.342.1117M} measured a median spectral index of $-0.83$ across 843--1400\,MHz by crossmatching SUMSS and NVSS. Cross-matching GLEAM with higher-resolution catalogues to investigate the low-frequency spectral behaviour of radio sources will be performed in the upcoming paper Line et al. (in prep).

\subsection{Source counts}\label{sec:source_counts}

Using selected subsections of the survey ($0^\mathrm{h}<\mathrm{RA}<3^\mathrm{h}$ and $-60\arcdeg<\mathrm{Dec}<-10\arcdeg$; $10^\mathrm{h}<\mathrm{RA}<12^\mathrm{h}$ and $-40\arcdeg<\mathrm{Dec}<-15\arcdeg$) where the noise is low and uniform ($6.8\pm1.3$\,mJy\,beam$^{-1}$), and the completeness is 90\,\% at 50\,mJy, we calculate the normalised Euclidean differential source counts with respect to source integrated flux density. These are tabulated in \Tab~\ref{tab:source_counts}.

\begin{table}
\caption{Normalised Euclidean differential source counts from the GLEAM 200\,MHz wideband catalogue.
\label{tab:source_counts}}
\begin{tabular}{ccccc}
\hline
$S_\mathrm{low}$ & $S_\mathrm{high}$ & $S_\mathrm{mid}$ & Raw count & $S^\frac{5}{2} \frac{dN}{dS}$ \tabularnewline
Jy & Jy & Jy &  & Jy$^\frac{3}{2}$ \tabularnewline
\hline
\hline
$0.023$ & $0.028$ & $0.025$ & 13 & $0.3118\pm0.0173$ \tabularnewline
$0.028$ & $0.035$ & $0.031$ & 302 & $10.12\pm0.117$ \tabularnewline
$0.035$ & $0.044$ & $0.039$ & 2361 & $110.6\pm0.455$ \tabularnewline
$0.044$ & $0.055$ & $0.049$ & 5060 & $331.3\pm0.931$ \tabularnewline
$0.055$ & $0.069$ & $0.061$ & 6200 & $567.3\pm1.44$ \tabularnewline
$0.069$ & $0.086$ & $0.077$ & 5673 & $725.4\pm1.93$ \tabularnewline
$0.086$ & $0.107$ & $0.096$ & 5105 & $912.3\pm2.55$ \tabularnewline
$0.107$ & $0.134$ & $0.120$ & 4295 & $1073\pm3.27$ \tabularnewline
$0.134$ & $0.168$ & $0.150$ & 3573 & $1247\pm4.17$ \tabularnewline
$0.168$ & $0.210$ & $0.188$ & 2991 & $1459\pm5.34$ \tabularnewline
$0.210$ & $0.262$ & $0.234$ & 2504 & $1707\pm6.82$ \tabularnewline
$0.262$ & $0.328$ & $0.293$ & 2013 & $1918\pm8.55$ \tabularnewline
$0.328$ & $0.410$ & $0.366$ & 1633 & $2174\pm10.8$ \tabularnewline
$0.410$ & $0.512$ & $0.458$ & 1342 & $2497\pm13.6$ \tabularnewline
$0.512$ & $0.640$ & $0.572$ & 1080 & $2808\pm17.1$ \tabularnewline
$0.640$ & $0.800$ & $0.716$ & 795 & $2889\pm20.5$ \tabularnewline
$0.800$ & $1.000$ & $0.894$ & 652 & $3311\pm25.9$ \tabularnewline
$1.000$ & $1.250$ & $1.118$ & 465 & $3301\pm30.6$ \tabularnewline
$1.250$ & $1.562$ & $1.398$ & 414 & $4107\pm40.4$ \tabularnewline
$1.563$ & $1.953$ & $1.747$ & 284 & $3937\pm46.7$ \tabularnewline
$1.953$ & $2.441$ & $2.184$ & 203 & $3933\pm55.2$ \tabularnewline
$2.441$ & $3.052$ & $2.730$ & 141 & $3818\pm64.3$ \tabularnewline
$3.052$ & $3.815$ & $3.412$ & 101 & $3822\pm76.1$ \tabularnewline
$3.815$ & $4.768$ & $4.265$ & 76 & $4019\pm92.2$ \tabularnewline
$4.768$ & $5.960$ & $5.331$ & 30 & $2217\pm 81$ \tabularnewline
$5.960$ & $7.451$ & $6.664$ & 27 & $2789\pm107$ \tabularnewline
$7.451$ & $9.313$ & $8.330$ & 26 & $3753\pm147$ \tabularnewline
$9.313$ & $11.642$ & $10.413$ & 18 & $3631\pm171$ \tabularnewline
$11.642$ & $14.552$ & $13.016$ & 12 & $3383\pm195$ \tabularnewline
$14.552$ & $18.190$ & $16.270$ & 7 & $2758\pm208$ \tabularnewline
$18.190$ & $22.737$ & $20.337$ & 7 & $3855\pm291$ \tabularnewline
$22.737$ & $28.422$ & $25.421$ & 1 & $769.6\pm154$ \tabularnewline
$28.422$ & $35.527$ & $31.776$ & 1 & $1076\pm215$ \tabularnewline
\hline
\end{tabular}
\end{table}

\begin{figure*}
	\centering
	\includegraphics[width=0.8\textwidth]{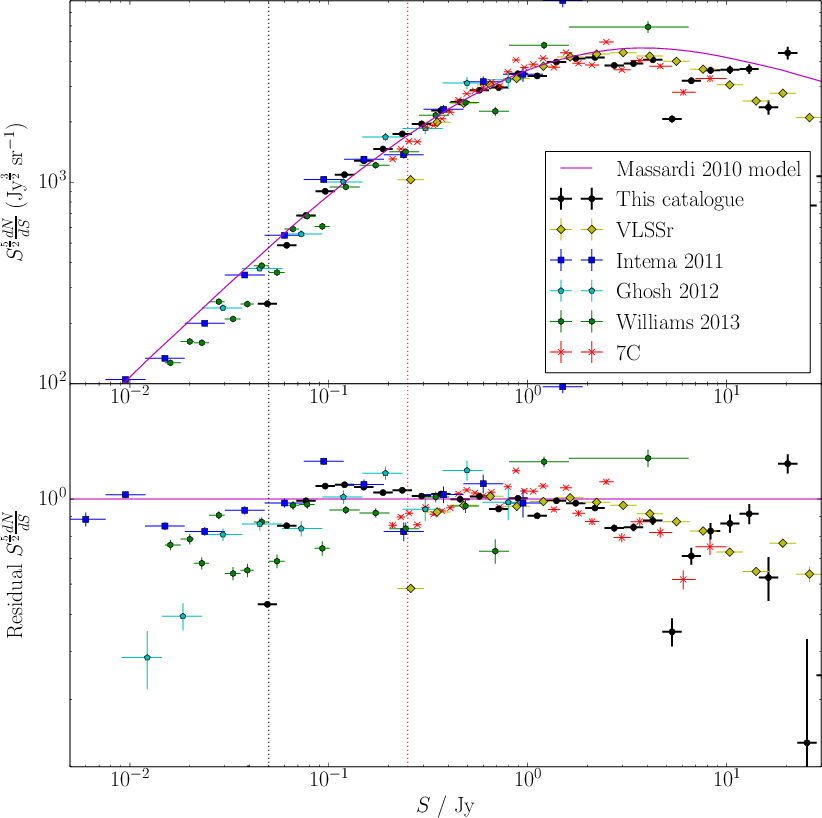} 
	\caption{Normalised Euclidean source counts at 150\,MHz for GLEAM (black circles), 7C (red crosses), VLSSr (yellow diamonds), and counts by \protect\cite{2011A+A...535A..38I} (blue squares), \protect\cite{2012MNRAS.426.3295G} (cyan pentagons), and \protect\cite{2013A+A...549A..55W} (green hexagons). The 150\,MHz astrophysical source evolution and luminosity model of \protect\cite{2010MNRAS.404..532M} is shown as a magenta line. The 90\,\% completeness flux density of GLEAM in this region is plotted as a 50\,mJy dotted line; the 7C completeness is plotted as a red dotted line. The top panel shows the source counts and the bottom panel shows the residuals after the \protect\cite{2010MNRAS.404..532M} model was subtracted.
    \label{fig:source_counts}}
\end{figure*}

We compare our source counts with those from the 7th Cambridge survey at 150\,MHz \citep[7C][]{2007MNRAS.382.1639H}, VLSSr at 74\,MHz, and those by \cite{2011A+A...535A..38I}, \cite{2012MNRAS.426.3295G}, and \cite{2013A+A...549A..55W}, whose measurements with the GMRT at 153\,MHz constrain the counts from 6--400\,mJy.
We apply a spectral power law scaling of $\alpha=-0.75$ to bring all surveys to a common frequency of 150\,MHz, and plot the surveys and the 150\,MHz astrophysical source evolution and luminosity model of \cite{2010MNRAS.404..532M} in \Fig~\ref{fig:source_counts}.

There is excellent agreement between all the surveys and the \cite{2010MNRAS.404..532M} model for the range 0.5--3\,Jy. Fainter and brighter than this, the surveys plotted disagree with the model; the bright end discrepancy could be due to the selection of fields without bright sources, which would otherwise increase the local RMS.

GLEAM is a relatively low-resolution survey, and confuses unrelated galaxies which are resolved by the other surveys. This may push many sources in faint bins into brighter bins, as they are confused together into larger, brighter sources by the PSF; this may account for the slight increase in observed counts around 100\,mJy. The completeness limit is also clearly evident at the low flux density end; this region of sky is 90\,\% complete at 50\,mJy, and at this flux density the GLEAM counts drop dramatically. \cite{2016MNRAS.459.3314F} correct for these effects and derive detailed source counts for the 600~square degrees imaged by \cite{2016MNRAS.458.1057O}. A future paper, Franzen et al (in prep) will perform the same analysis for the GLEAM survey, over the full bandwidth of 72--231\,MHz.

\subsection{Spectral fitting with GLEAM data}

The spectral coverage of the GLEAM survey represents a divergence from past radio surveys, which mostly surveyed the radio sky at one frequency with a small bandwidth (e.g. 3C, MRC, SUMSS, NVSS, VLSSr etc). The twenty flux density measurements between 72 and 231\,MHz reported in GLEAM provides an unparalleled data set for spectral analysis of radio sources. However, with this advancement in bandwidth come statistical challenges for spectral modelling and correctly combining data from many different telescopes.

For example, since self-calibration and multi-frequency synthesis were performed on the full 30.72\,MHz observing bandwidth, before it is spilt into four narrower sub-bands of 7.68\,MHz, the four derived subband flux densities within one band are highly correlated. Classical and sidelobe confusion also produce correlated noise that is dependent on the flux density of the source \citep{1973ApJ...183....1M,2012ApJ...758...23C}, with faint sources ($<$1\,Jy) having a more significant degree of correlation across the entire MWA band, compared to bright sources. Additionally, due to the correction of the primary beam uncertainties, all of the flux density measurements are correlated. The combination of these effects generates a complex covariance function that should be taken into account when combining GLEAM data with that from other radio telescopes or surveys. If this correlation between the GLEAM data points are not taken into account, the remaining trends present in the GLEAM flux density measurements can dominate any physical relations. Most notably, the 8\,\% flux scale error (\Sect~\ref{sec:systematic-error}) \textit{does not} reduce by $\frac{1}{\sqrt{20}}$ when the twenty sub-band data points are combined (either in a fit or a weighted average), so the points should never be treated as independent data points when comparing to data outside the GLEAM survey.

\begin{figure}
	\centering
	\includegraphics[width=0.45\textwidth]{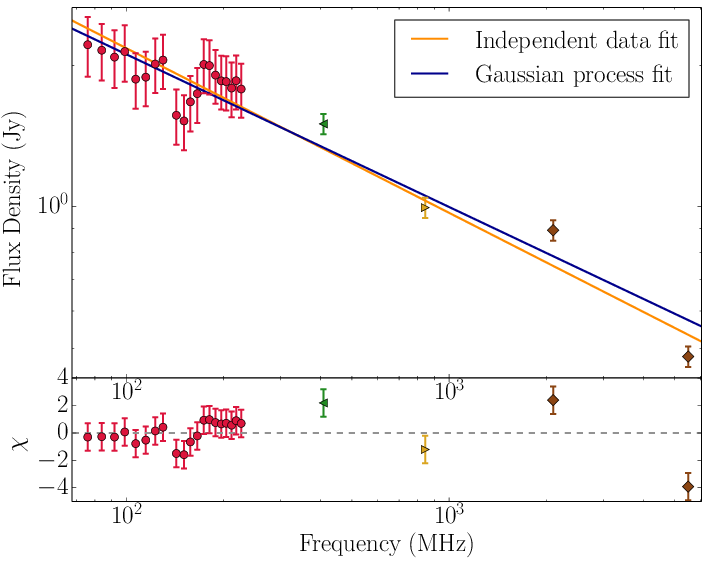} 
	\caption{Spectral energy distribution of PKS~B2059-786 highlighting the different model fits, assuming the flux density measurements are correlated or independent. Red circles, green leftward-facing triangle, yellow rightward-facing triangle, and brown diamonds represent data points from GLEAM, MRC, SUMSS, and the Australia Telescope Compact Array calibrator database, respectively. The power-law model fit assuming all the data points are independent is shown in dark orange. The model fit assuming the covariance matrix of the MWA flux density measurements is described by a blocked Mat\'{e}rn covariance function is shown in blue. The $\chi$-values for the Gaussian process model fit to the data are displayed in the panel below the spectral energy distribution.
    \label{fig:gp_fit}}
\end{figure}

While it is currently not possible to calculate the exact form of the covariance function that exists between the GLEAM flux density measurements, an approximation can be made using Gaussian processes. For example, the correlation between the GLEAM flux density measurements is found to be accurately described by a blocked Mat\'{e}rn covariance function \citep{Rasmussen2006}, which produces a stronger correlation between flux density measurements close in frequency space than further away. An example is provided in Figure \ref{fig:gp_fit}, which demonstrates how the fit is incorrect if one assumes the GLEAM data points are independent.

\section{Conclusions}
\label{sec:conclusions}

We have performed a large-scale radio sky survey that we estimate is 99.5\,\% complete at 1\,Jy for the sky south of Dec~$+30\arcdeg$, excluding the Galactic Plane and a small number of other regions. The estimated reliability is \pctreliable\,\%. The completeness varies over the sky, and we provide machine-readable maps of the completeness at different flux density levels to assist the catalogue user.
Using a deep, wideband image formed across 170--231\,MHz, we measured the flux density of \nsrc detected sources at 20 frequencies spanning 72--231\,MHz. Source spectral indices derived across this bandwidth agree with results from other experiments using much wider frequency lever-arms.

The overall flux density scale accuracy is estimated to be 8\,\% for 90\,\% of the surveyed area. This survey is on the \cite{Baars1977} flux density scale; a future paper will examine the agreement between GLEAM and MSSS, unifying the low-frequency flux scales over the whole sky. This catalogue makes possible reliable flux calibration of other low-frequency southern sky experiments, such as the search for the Epoch of Reionisation by the MWA, PAPER, and eventually the low-frequency Square Kilometre Array.

The low-frequency flux densities and spectral indices of hundreds of thousands of radio galaxies are now available, and cross-matching with higher-resolution data to disentangle confused pairs and reveal morphology should maximise the utility of our low-frequency flux density measurements. In this large dataset, there are also likely to be some interesting serendipitous detections. Future papers will search the survey for transient sources, reprocess the data with a weighting scheme which increases the impact of the short baselines of the array, thereby highlighting more diffuse structures, and publish the Galactic Plane and Magellanic Clouds.

In addition to the observations used to create the catalogue presented here, the sky has also been observed twice more during the second year of observations. Processing of these new data will reduce the overall noise of the survey, where not already dominated by sidelobe confusion, and improve the sky coverage over the first year of observations. More advanced processing techniques such as the application of direction-dependent gains, may increase the survey depth further.

All data (images and catalogue) are publicly available at the MWA Telescope website on the World Wide Web, \texttt{http://www.mwatelescope.org}.
\section*{Acknowledgements}
J.~R.~C. thanks Richard Perley of NRAO for providing expertise in reducing the P-band VLA data.
We acknowledge the International Centre for Radio Astronomy Research (ICRAR), a Joint Venture of Curtin University and The University of Western Australia, funded by the Western Australian State government.
This scientific work makes use of the Murchison Radio-astronomy Observatory, operated by CSIRO. We acknowledge the Wajarri Yamatji people as the traditional owners of the Observatory site. Support for the operation of the MWA is provided by the Australian Government (NCRIS), under a contract to Curtin University administered by Astronomy Australia Limited. We acknowledge the Pawsey Supercomputing Centre which is supported by the Western Australian and Australian Governments.
This research was undertaken with the assistance of resources from the National Computational Infrastructure (NCI), which is supported by the Australian Government.
This research has made use of the National Aeronautics and Space Administration (NASA) / Infrared Processing and Analysis Center (IPAC) Infrared Science Archive and the NASA/IPAC Extragalactic Database (NED) which are operated by the Jet Propulsion Laboratory, California Institute of Technology, under contract with NASA. This research has also made use of NASA's Astrophysics Data System.
\bibliographystyle{mn2e_columbia}

\setlength{\labelwidth}{0pt}

\bibliography{refs}

\appendix
\section{Appendix: List of column headings in the catalogue}\label{sec:appendix}
%
%
\onecolumn
Column numbers, names, and units for the catalogue.
Source names follow International Astronomical Union naming conventions for co-ordinate-based naming.
Background and RMS measurements were performed by \textsc{BANE};
PSF measurements were peformed using in-house software as described in \Sect~\ref{sec:ionopsf};
the absolute error on the flux density scale was derived as described in \Sect~\ref{sec:corr_pb};
the fitted spectral index parameters were derived as described in \Sect~\ref{sec:alpha};
all other measurements were made using \textsc{Aegean}. \textsc{Aegean} incorporates a constrained fitting algorithm.
Shape parameters with an error of $-1$ indicate that the reported value is equal to either the upper or lower fitting constraint.
The columns with the subscript ``wide'' are derived from the 200\,MHz wide-band image.
Subsequently, the subscript indicates the central frequency of the measurement, in MHz.
These sub-band measurements are made using the priorised fitting mode of Aegean, where the position and shape of the source are determined from the wide-band image, and only the flux density is fitted (see \Sect~\ref{sec:priorised}).
Note therefore that some columns in the priorised fit do not have error bars, because they are linearly propagated from the wideband image values (e.g. major axis $a$).
\label{tab:appendix}
\begin{longtable}{cccc}
\hline
Number & Name & Unit & Description \tabularnewline
\hline
1 & Name & hh:mm:ss+dd:mm:ss & International Astronomical Union name  \tabularnewline 
2 & background\_wide & Jy\,beam$^{-1}$ & Background in wideband image \tabularnewline 
3 & local\_rms\_wide & Jy\,beam$^{-1}$ & Local RMS in wideband image \tabularnewline 
4 & ra\_str & hh:mm:ss & Right ascension \tabularnewline 
5 & dec\_str & dd:mm:ss & Declination \tabularnewline 
6 & RAJ2000 & $\arcdeg$ & Right ascension \tabularnewline 
7 & err\_RAJ2000 & $\arcdeg$ & Error on RA \tabularnewline 
8 & DEJ2000 & $\arcdeg$ & Declination \tabularnewline 
9 & err\_DEJ2000 & $\arcdeg$ & Error on Dec \tabularnewline 
10 & peak\_flux\_wide & Jy\,beam$^{-1}$ & Peak flux density in wideband image \tabularnewline 
11 & err\_peak\_flux\_wide & Jy\,beam$^{-1}$ & Fitting error on peak flux density in wideband image \tabularnewline 
12 & int\_flux\_wide & Jy & Integrated flux density in wideband image \tabularnewline 
13 & err\_int\_flux\_wide & Jy & Error on integrated flux density in wideband image \tabularnewline 
14 & a\_wide & $\arcsec$ & Major axis of source in wideband image \tabularnewline 
15 & err\_a\_wide & $\arcsec$ & Error on major axis of source in wideband image \tabularnewline 
16 & b\_wide & $\arcsec$ & Minor axis of source in wideband image \tabularnewline 
17 & err\_b\_wide & $\arcsec$ & Error on minor axis of source in wideband image \tabularnewline 
18 & pa\_wide & $\arcdeg$ & Postion angle of source in wideband image \tabularnewline 
19 & err\_pa\_wide & $\arcdeg$ & Error on position angle of source in wideband image \tabularnewline 
20 & residual\_mean\_wide & Jy\,beam$^{-1}$ & Mean of residual after source fitting in wideband image \tabularnewline 
21 & residual\_std\_wide & Jy\,beam$^{-1}$ & Standard deviation of residual after source fitting \tabularnewline 
22 & err\_abs\_flux\_pct & \% & Percent error in absolute flux scale - all frequenciesc \tabularnewline 
23 & err\_fit\_flux\_pct & \% & Percent error on internal flux scale - all frequencies \tabularnewline
24 & psf\_a\_wide & $\arcsec$ & Major axis of PSF at location of source in wideband image \tabularnewline
25 & psf\_b\_wide & $\arcsec$ & Minor axis of PSF at location of source in wideband image \tabularnewline
26 & psf\_pa\_wide & $\arcdeg$ & Position angle of PSF at location of source in wideband image \tabularnewline
27 & background\_076 & Jy\,beam$^{-1}$ & Background at 76\,MHz \tabularnewline
28 & local\_rms\_076 & Jy\,beam$^{-1}$ & Local RMS at 76\,MHz \tabularnewline
29 & peak\_flux\_076 & Jy\,beam$^{-1}$ & Peak flux density at 76\,MHz \tabularnewline
30 & err\_peak\_flux\_076 & Jy\,beam$^{-1}$ & Fitting error on peak flux density at 76\,MHz \tabularnewline
31 & int\_flux\_076 & Jy & Integrated flux density at 76\,MHz \tabularnewline
32 & err\_int\_flux\_076 & Jy & Fitting error on integrated flux density at 76\,MHz \tabularnewline
33 & a\_076 & $\arcsec$ & Major axis of source at 76\,MHz \tabularnewline
34 & b\_076 & $\arcsec$ & Minor axis of source at 76\,MHz \tabularnewline
35 & pa\_076 & $\arcdeg$ & Position angle of source at 76\,MHz \tabularnewline
36 & residual\_mean\_076 & Jy\,beam$^{-1}$ & Mean of residual after source fitting at 76\,MHz \tabularnewline
37 & residual\_std\_076 & Jy\,beam$^{-1}$ & Standard deviation of residual after source fitting at 76\,MHz \tabularnewline
38 & psf\_a\_076 & $\arcsec$ & Major axis of PSF at location of source at 76\,MHz \tabularnewline
39 & psf\_b\_076 & $\arcsec$ & Minor axis of PSF at location of source at 76\,MHz \tabularnewline
40 & psf\_pa\_076 & $\arcdeg$ & Position angle of PSF at location of source at 76\,MHz \tabularnewline
41 & background\_084 & Jy\,beam$^{-1}$ & Background at 84\,MHz \tabularnewline
42 & local\_rms\_084 & Jy\,beam$^{-1}$ & Local RMS at 84\,MHz \tabularnewline
43 & peak\_flux\_084 & Jy\,beam$^{-1}$ & Peak flux density at 84\,MHz \tabularnewline
44 & err\_peak\_flux\_084 & Jy\,beam$^{-1}$ & Fitting error on peak flux density at 84\,MHz \tabularnewline
45 & int\_flux\_084 & Jy & Integrated flux density at 84\,MHz \tabularnewline
46 & err\_int\_flux\_084 & Jy & Fitting error on integrated flux density at 84\,MHz \tabularnewline
47 & a\_084 & $\arcsec$ & Major axis of source at 84\,MHz \tabularnewline
48 & b\_084 & $\arcsec$ & Minor axis of source at 84\,MHz \tabularnewline
49 & pa\_084 & $\arcdeg$ & Position angle of source at 84\,MHz \tabularnewline
50 & residual\_mean\_084 & Jy\,beam$^{-1}$ & Mean of residual after source fitting at 84\,MHz \tabularnewline
51 & residual\_std\_084 & Jy\,beam$^{-1}$ & Standard deviation of residual after source fitting at 84\,MHz \tabularnewline
52 & psf\_a\_084 & $\arcsec$ & Major axis of PSF at location of source at 84\,MHz \tabularnewline
53 & psf\_b\_084 & $\arcsec$ & Minor axis of PSF at location of source at 84\,MHz \tabularnewline
54 & psf\_pa\_084 & $\arcdeg$ & Position angle of PSF at location of source at 84\,MHz \tabularnewline
55 & background\_092 & Jy\,beam$^{-1}$ & Background at 92\,MHz \tabularnewline
56 & local\_rms\_092 & Jy\,beam$^{-1}$ & Local RMS at 92\,MHz \tabularnewline
57 & peak\_flux\_092 & Jy\,beam$^{-1}$ & Peak flux density at 92\,MHz \tabularnewline
58 & err\_peak\_flux\_092 & Jy\,beam$^{-1}$ & Fitting error on peak flux density at 92\,MHz \tabularnewline
59 & int\_flux\_092 & Jy & Integrated flux density at 92\,MHz \tabularnewline
60 & err\_int\_flux\_092 & Jy & Fitting error on integrated flux density at 92\,MHz \tabularnewline
61 & a\_092 & $\arcsec$ & Major axis of source at 92\,MHz \tabularnewline
62 & b\_092 & $\arcsec$ & Minor axis of source at 92\,MHz \tabularnewline
63 & pa\_092 & $\arcdeg$ & Position angle of source at 92\,MHz \tabularnewline
64 & residual\_mean\_092 & Jy\,beam$^{-1}$ & Mean of residual after source fitting at 92\,MHz \tabularnewline
65 & residual\_std\_092 & Jy\,beam$^{-1}$ & Standard deviation of residual after source fitting at 92\,MHz \tabularnewline
66 & psf\_a\_092 & $\arcsec$ & Major axis of PSF at location of source at 92\,MHz \tabularnewline
67 & psf\_b\_092 & $\arcsec$ & Minor axis of PSF at location of source at 92\,MHz \tabularnewline
68 & psf\_pa\_092 & $\arcdeg$ & Position angle of PSF at location of source at 92\,MHz \tabularnewline
69 & background\_099 & Jy\,beam$^{-1}$ & Background at 99\,MHz \tabularnewline
70 & local\_rms\_099 & Jy\,beam$^{-1}$ & Local RMS at 99\,MHz \tabularnewline
71 & peak\_flux\_099 & Jy\,beam$^{-1}$ & Peak flux density at 99\,MHz \tabularnewline
72 & err\_peak\_flux\_099 & Jy\,beam$^{-1}$ & Fitting error on peak flux density at 99\,MHz \tabularnewline
73 & int\_flux\_099 & Jy & Integrated flux density at 99\,MHz \tabularnewline
74 & err\_int\_flux\_099 & Jy & Fitting error on integrated flux density at 99\,MHz \tabularnewline
75 & a\_099 & $\arcsec$ & Major axis of source at 99\,MHz \tabularnewline
76 & b\_099 & $\arcsec$ & Minor axis of source at 99\,MHz \tabularnewline
77 & pa\_099 & $\arcdeg$ & Position angle of source at 99\,MHz \tabularnewline
78 & residual\_mean\_099 & Jy\,beam$^{-1}$ & Mean of residual after source fitting at 99\,MHz \tabularnewline
79 & residual\_std\_099 & Jy\,beam$^{-1}$ & Standard deviation of residual after source fitting at 99\,MHz \tabularnewline
80 & psf\_a\_099 & $\arcsec$ & Major axis of PSF at location of source at 99\,MHz \tabularnewline
81 & psf\_b\_099 & $\arcsec$ & Minor axis of PSF at location of source at 99\,MHz \tabularnewline
82 & psf\_pa\_099 & $\arcdeg$ & Position angle of PSF at location of source at 99\,MHz \tabularnewline
83 & background\_107 & Jy\,beam$^{-1}$ & Background at 107\,MHz \tabularnewline
84 & local\_rms\_107 & Jy\,beam$^{-1}$ & Local RMS at 107\,MHz \tabularnewline
85 & peak\_flux\_107 & Jy\,beam$^{-1}$ & Peak flux density at 107\,MHz \tabularnewline
86 & err\_peak\_flux\_107 & Jy\,beam$^{-1}$ & Fitting error on peak flux density at 107\,MHz \tabularnewline
87 & int\_flux\_107 & Jy & Integrated flux density at 107\,MHz \tabularnewline
88 & err\_int\_flux\_107 & Jy & Fitting error on integrated flux density at 107\,MHz \tabularnewline
89 & a\_107 & $\arcsec$ & Major axis of source at 107\,MHz \tabularnewline
90 & b\_107 & $\arcsec$ & Minor axis of source at 107\,MHz \tabularnewline
91 & pa\_107 & $\arcdeg$ & Position angle of source at 107\,MHz \tabularnewline
92 & residual\_mean\_107 & Jy\,beam$^{-1}$ & Mean of residual after source fitting at 107\,MHz \tabularnewline
93 & residual\_std\_107 & Jy\,beam$^{-1}$ & Standard deviation of residual after source fitting at 107\,MHz \tabularnewline
94 & psf\_a\_107 & $\arcsec$ & Major axis of PSF at location of source at 107\,MHz \tabularnewline
95 & psf\_b\_107 & $\arcsec$ & Minor axis of PSF at location of source at 107\,MHz \tabularnewline
96 & psf\_pa\_107 & $\arcdeg$ & Position angle of PSF at location of source at 107\,MHz \tabularnewline
97 & background\_115 & Jy\,beam$^{-1}$ & Background at 115\,MHz \tabularnewline
98 & local\_rms\_115 & Jy\,beam$^{-1}$ & Local RMS at 115\,MHz \tabularnewline
99 & peak\_flux\_115 & Jy\,beam$^{-1}$ & Peak flux density at 115\,MHz \tabularnewline
100 & err\_peak\_flux\_115 & Jy\,beam$^{-1}$ & Fitting error on peak flux density at 115\,MHz \tabularnewline
101 & int\_flux\_115 & Jy & Integrated flux density at 115\,MHz \tabularnewline
102 & err\_int\_flux\_115 & Jy & Fitting error on integrated flux density at 115\,MHz \tabularnewline
103 & a\_115 & $\arcsec$ & Major axis of source at 115\,MHz \tabularnewline
104 & b\_115 & $\arcsec$ & Minor axis of source at 115\,MHz \tabularnewline
105 & pa\_115 & $\arcdeg$ & Position angle of source at 115\,MHz \tabularnewline
106 & residual\_mean\_115 & Jy\,beam$^{-1}$ & Mean of residual after source fitting at 115\,MHz \tabularnewline
107 & residual\_std\_115 & Jy\,beam$^{-1}$ & Standard deviation of residual after source fitting at 115\,MHz \tabularnewline
108 & psf\_a\_115 & $\arcsec$ & Major axis of PSF at location of source at 115\,MHz \tabularnewline
109 & psf\_b\_115 & $\arcsec$ & Minor axis of PSF at location of source at 115\,MHz \tabularnewline
110 & psf\_pa\_115 & $\arcdeg$ & Position angle of PSF at location of source at 115\,MHz \tabularnewline
111 & background\_122 & Jy\,beam$^{-1}$ & Background at 122\,MHz \tabularnewline
112 & local\_rms\_122 & Jy\,beam$^{-1}$ & Local RMS at 122\,MHz \tabularnewline
113 & peak\_flux\_122 & Jy\,beam$^{-1}$ & Peak flux density at 122\,MHz \tabularnewline
114 & err\_peak\_flux\_122 & Jy\,beam$^{-1}$ & Fitting error on peak flux density at 122\,MHz \tabularnewline
115 & int\_flux\_122 & Jy & Integrated flux density at 122\,MHz \tabularnewline
116 & err\_int\_flux\_122 & Jy & Fitting error on integrated flux density at 122\,MHz \tabularnewline
117 & a\_122 & $\arcsec$ & Major axis of source at 122\,MHz \tabularnewline
118 & b\_122 & $\arcsec$ & Minor axis of source at 122\,MHz \tabularnewline
119 & pa\_122 & $\arcdeg$ & Position angle of source at 122\,MHz \tabularnewline
120 & residual\_mean\_122 & Jy\,beam$^{-1}$ & Mean of residual after source fitting at 122\,MHz \tabularnewline
121 & residual\_std\_122 & Jy\,beam$^{-1}$ & Standard deviation of residual after source fitting at 122\,MHz \tabularnewline
122 & psf\_a\_122 & $\arcsec$ & Major axis of PSF at location of source at 122\,MHz \tabularnewline
123 & psf\_b\_122 & $\arcsec$ & Minor axis of PSF at location of source at 122\,MHz \tabularnewline
124 & psf\_pa\_122 & $\arcdeg$ & Position angle of PSF at location of source at 122\,MHz \tabularnewline
125 & background\_130 & Jy\,beam$^{-1}$ & Background at 130\,MHz \tabularnewline
126 & local\_rms\_130 & Jy\,beam$^{-1}$ & Local RMS at 130\,MHz \tabularnewline
127 & peak\_flux\_130 & Jy\,beam$^{-1}$ & Peak flux density at 130\,MHz \tabularnewline
128 & err\_peak\_flux\_130 & Jy\,beam$^{-1}$ & Fitting error on peak flux density at 130\,MHz \tabularnewline
129 & int\_flux\_130 & Jy & Integrated flux density at 130\,MHz \tabularnewline
130 & err\_int\_flux\_130 & Jy & Fitting error on integrated flux density at 130\,MHz \tabularnewline
131 & a\_130 & $\arcsec$ & Major axis of source at 130\,MHz \tabularnewline
132 & b\_130 & $\arcsec$ & Minor axis of source at 130\,MHz \tabularnewline
133 & pa\_130 & $\arcdeg$ & Position angle of source at 130\,MHz \tabularnewline
134 & residual\_mean\_130 & Jy\,beam$^{-1}$ & Mean of residual after source fitting at 130\,MHz \tabularnewline
135 & residual\_std\_130 & Jy\,beam$^{-1}$ & Standard deviation of residual after source fitting at 130\,MHz \tabularnewline
136 & psf\_a\_130 & $\arcsec$ & Major axis of PSF at location of source at 130\,MHz \tabularnewline
137 & psf\_b\_130 & $\arcsec$ & Minor axis of PSF at location of source at 130\,MHz \tabularnewline
138 & psf\_pa\_130 & $\arcdeg$ & Position angle of PSF at location of source at 130\,MHz \tabularnewline
139 & background\_143 & Jy\,beam$^{-1}$ & Background at 143\,MHz \tabularnewline
140 & local\_rms\_143 & Jy\,beam$^{-1}$ & Local RMS at 143\,MHz \tabularnewline
141 & peak\_flux\_143 & Jy\,beam$^{-1}$ & Peak flux density at 143\,MHz \tabularnewline
142 & err\_peak\_flux\_143 & Jy\,beam$^{-1}$ & Fitting error on peak flux density at 143\,MHz \tabularnewline
143 & int\_flux\_143 & Jy & Integrated flux density at 143\,MHz \tabularnewline
144 & err\_int\_flux\_143 & Jy & Fitting error on integrated flux density at 143\,MHz \tabularnewline
145 & a\_143 & $\arcsec$ & Major axis of source at 143\,MHz \tabularnewline
146 & b\_143 & $\arcsec$ & Minor axis of source at 143\,MHz \tabularnewline
147 & pa\_143 & $\arcdeg$ & Position angle of source at 143\,MHz \tabularnewline
148 & residual\_mean\_143 & Jy\,beam$^{-1}$ & Mean of residual after source fitting at 143\,MHz \tabularnewline
149 & residual\_std\_143 & Jy\,beam$^{-1}$ & Standard deviation of residual after source fitting at 143\,MHz \tabularnewline
150 & psf\_a\_143 & $\arcsec$ & Major axis of PSF at location of source at 143\,MHz \tabularnewline
151 & psf\_b\_143 & $\arcsec$ & Minor axis of PSF at location of source at 143\,MHz \tabularnewline
152 & psf\_pa\_143 & $\arcdeg$ & Position angle of PSF at location of source at 143\,MHz \tabularnewline
153 & background\_151 & Jy\,beam$^{-1}$ & Background at 151\,MHz \tabularnewline
154 & local\_rms\_151 & Jy\,beam$^{-1}$ & Local RMS at 151\,MHz \tabularnewline
155 & peak\_flux\_151 & Jy\,beam$^{-1}$ & Peak flux density at 151\,MHz \tabularnewline
156 & err\_peak\_flux\_151 & Jy\,beam$^{-1}$ & Fitting error on peak flux density at 151\,MHz \tabularnewline
157 & int\_flux\_151 & Jy & Integrated flux density at 151\,MHz \tabularnewline
158 & err\_int\_flux\_151 & Jy & Fitting error on integrated flux density at 151\,MHz \tabularnewline
159 & a\_151 & $\arcsec$ & Major axis of source at 151\,MHz \tabularnewline
160 & b\_151 & $\arcsec$ & Minor axis of source at 151\,MHz \tabularnewline
161 & pa\_151 & $\arcdeg$ & Position angle of source at 151\,MHz \tabularnewline
162 & residual\_mean\_151 & Jy\,beam$^{-1}$ & Mean of residual after source fitting at 151\,MHz \tabularnewline
163 & residual\_std\_151 & Jy\,beam$^{-1}$ & Standard deviation of residual after source fitting at 151\,MHz \tabularnewline
164 & psf\_a\_151 & $\arcsec$ & Major axis of PSF at location of source at 151\,MHz \tabularnewline
165 & psf\_b\_151 & $\arcsec$ & Minor axis of PSF at location of source at 151\,MHz \tabularnewline
166 & psf\_pa\_151 & $\arcdeg$ & Position angle of PSF at location of source at 151\,MHz \tabularnewline
167 & background\_158 & Jy\,beam$^{-1}$ & Background at 158\,MHz \tabularnewline
168 & local\_rms\_158 & Jy\,beam$^{-1}$ & Local RMS at 158\,MHz \tabularnewline
169 & peak\_flux\_158 & Jy\,beam$^{-1}$ & Peak flux density at 158\,MHz \tabularnewline
170 & err\_peak\_flux\_158 & Jy\,beam$^{-1}$ & Fitting error on peak flux density at 158\,MHz \tabularnewline
171 & int\_flux\_158 & Jy & Integrated flux density at 158\,MHz \tabularnewline
172 & err\_int\_flux\_158 & Jy & Fitting error on integrated flux density at 158\,MHz \tabularnewline
173 & a\_158 & $\arcsec$ & Major axis of source at 158\,MHz \tabularnewline
174 & b\_158 & $\arcsec$ & Minor axis of source at 158\,MHz \tabularnewline
175 & pa\_158 & $\arcdeg$ & Position angle of source at 158\,MHz \tabularnewline
176 & residual\_mean\_158 & Jy\,beam$^{-1}$ & Mean of residual after source fitting at 158\,MHz \tabularnewline
177 & residual\_std\_158 & Jy\,beam$^{-1}$ & Standard deviation of residual after source fitting at 158\,MHz \tabularnewline
178 & psf\_a\_158 & $\arcsec$ & Major axis of PSF at location of source at 158\,MHz \tabularnewline
179 & psf\_b\_158 & $\arcsec$ & Minor axis of PSF at location of source at 158\,MHz \tabularnewline
180 & psf\_pa\_158 & $\arcdeg$ & Position angle of PSF at location of source at 158\,MHz \tabularnewline
181 & background\_166 & Jy\,beam$^{-1}$ & Background at 166\,MHz \tabularnewline
182 & local\_rms\_166 & Jy\,beam$^{-1}$ & Local RMS at 166\,MHz \tabularnewline
183 & peak\_flux\_166 & Jy\,beam$^{-1}$ & Peak flux density at 166\,MHz \tabularnewline
184 & err\_peak\_flux\_166 & Jy\,beam$^{-1}$ & Fitting error on peak flux density at 166\,MHz \tabularnewline
185 & int\_flux\_166 & Jy & Integrated flux density at 166\,MHz \tabularnewline
186 & err\_int\_flux\_166 & Jy & Fitting error on integrated flux density at 166\,MHz \tabularnewline
187 & a\_166 & $\arcsec$ & Major axis of source at 166\,MHz \tabularnewline
188 & b\_166 & $\arcsec$ & Minor axis of source at 166\,MHz \tabularnewline
189 & pa\_166 & $\arcdeg$ & Position angle of source at 166\,MHz \tabularnewline
190 & residual\_mean\_166 & Jy\,beam$^{-1}$ & Mean of residual after source fitting at 166\,MHz \tabularnewline
191 & residual\_std\_166 & Jy\,beam$^{-1}$ & Standard deviation of residual after source fitting at 166\,MHz \tabularnewline
192 & psf\_a\_166 & $\arcsec$ & Major axis of PSF at location of source at 166\,MHz \tabularnewline
193 & psf\_b\_166 & $\arcsec$ & Minor axis of PSF at location of source at 166\,MHz \tabularnewline
194 & psf\_pa\_166 & $\arcdeg$ & Position angle of PSF at location of source at 166\,MHz \tabularnewline
195 & background\_174 & Jy\,beam$^{-1}$ & Background at 174\,MHz \tabularnewline
196 & local\_rms\_174 & Jy\,beam$^{-1}$ & Local RMS at 174\,MHz \tabularnewline
197 & peak\_flux\_174 & Jy\,beam$^{-1}$ & Peak flux density at 174\,MHz \tabularnewline
198 & err\_peak\_flux\_174 & Jy\,beam$^{-1}$ & Fitting error on peak flux density at 174\,MHz \tabularnewline
199 & int\_flux\_174 & Jy & Integrated flux density at 174\,MHz \tabularnewline
200 & err\_int\_flux\_174 & Jy & Fitting error on integrated flux density at 174\,MHz \tabularnewline
201 & a\_174 & $\arcsec$ & Major axis of source at 174\,MHz \tabularnewline
202 & b\_174 & $\arcsec$ & Minor axis of source at 174\,MHz \tabularnewline
203 & pa\_174 & $\arcdeg$ & Position angle of source at 174\,MHz \tabularnewline
204 & residual\_mean\_174 & Jy\,beam$^{-1}$ & Mean of residual after source fitting at 174\,MHz \tabularnewline
205 & residual\_std\_174 & Jy\,beam$^{-1}$ & Standard deviation of residual after source fitting at 174\,MHz \tabularnewline
206 & psf\_a\_174 & $\arcsec$ & Major axis of PSF at location of source at 174\,MHz \tabularnewline
207 & psf\_b\_174 & $\arcsec$ & Minor axis of PSF at location of source at 174\,MHz \tabularnewline
208 & psf\_pa\_174 & $\arcdeg$ & Position angle of PSF at location of source at 174\,MHz \tabularnewline
209 & background\_181 & Jy\,beam$^{-1}$ & Background at 181\,MHz \tabularnewline
210 & local\_rms\_181 & Jy\,beam$^{-1}$ & Local RMS at 181\,MHz \tabularnewline
211 & peak\_flux\_181 & Jy\,beam$^{-1}$ & Peak flux density at 181\,MHz \tabularnewline
212 & err\_peak\_flux\_181 & Jy\,beam$^{-1}$ & Fitting error on peak flux density at 181\,MHz \tabularnewline
213 & int\_flux\_181 & Jy & Integrated flux density at 181\,MHz \tabularnewline
214 & err\_int\_flux\_181 & Jy & Fitting error on integrated flux density at 181\,MHz \tabularnewline
215 & a\_181 & $\arcsec$ & Major axis of source at 181\,MHz \tabularnewline
216 & b\_181 & $\arcsec$ & Minor axis of source at 181\,MHz \tabularnewline
217 & pa\_181 & $\arcdeg$ & Position angle of source at 181\,MHz \tabularnewline
218 & residual\_mean\_181 & Jy\,beam$^{-1}$ & Mean of residual after source fitting at 181\,MHz \tabularnewline
219 & residual\_std\_181 & Jy\,beam$^{-1}$ & Standard deviation of residual after source fitting at 181\,MHz \tabularnewline
220 & psf\_a\_181 & $\arcsec$ & Major axis of PSF at location of source at 181\,MHz \tabularnewline
221 & psf\_b\_181 & $\arcsec$ & Minor axis of PSF at location of source at 181\,MHz \tabularnewline
222 & psf\_pa\_181 & $\arcdeg$ & Position angle of PSF at location of source at 181\,MHz \tabularnewline
223 & background\_189 & Jy\,beam$^{-1}$ & Background at 189\,MHz \tabularnewline
224 & local\_rms\_189 & Jy\,beam$^{-1}$ & Local RMS at 189\,MHz \tabularnewline
225 & peak\_flux\_189 & Jy\,beam$^{-1}$ & Peak flux density at 189\,MHz \tabularnewline
226 & err\_peak\_flux\_189 & Jy\,beam$^{-1}$ & Fitting error on peak flux density at 189\,MHz \tabularnewline
227 & int\_flux\_189 & Jy & Integrated flux density at 189\,MHz \tabularnewline
228 & err\_int\_flux\_189 & Jy & Fitting error on integrated flux density at 189\,MHz \tabularnewline
229 & a\_189 & $\arcsec$ & Major axis of source at 189\,MHz \tabularnewline
230 & b\_189 & $\arcsec$ & Minor axis of source at 189\,MHz \tabularnewline
231 & pa\_189 & $\arcdeg$ & Position angle of source at 189\,MHz \tabularnewline
232 & residual\_mean\_189 & Jy\,beam$^{-1}$ & Mean of residual after source fitting at 189\,MHz \tabularnewline
233 & residual\_std\_189 & Jy\,beam$^{-1}$ & Standard deviation of residual after source fitting at 189\,MHz \tabularnewline
234 & psf\_a\_189 & $\arcsec$ & Major axis of PSF at location of source at 189\,MHz \tabularnewline
235 & psf\_b\_189 & $\arcsec$ & Minor axis of PSF at location of source at 189\,MHz \tabularnewline
236 & psf\_pa\_189 & $\arcdeg$ & Position angle of PSF at location of source at 189\,MHz \tabularnewline
237 & background\_197 & Jy\,beam$^{-1}$ & Background at 197\,MHz \tabularnewline
238 & local\_rms\_197 & Jy\,beam$^{-1}$ & Local RMS at 197\,MHz \tabularnewline
239 & peak\_flux\_197 & Jy\,beam$^{-1}$ & Peak flux density at 197\,MHz \tabularnewline
240 & err\_peak\_flux\_197 & Jy\,beam$^{-1}$ & Fitting error on peak flux density at 197\,MHz \tabularnewline
241 & int\_flux\_197 & Jy & Integrated flux density at 197\,MHz \tabularnewline
242 & err\_int\_flux\_197 & Jy & Fitting error on integrated flux density at 197\,MHz \tabularnewline
243 & a\_197 & $\arcsec$ & Major axis of source at 197\,MHz \tabularnewline
244 & b\_197 & $\arcsec$ & Minor axis of source at 197\,MHz \tabularnewline
245 & pa\_197 & $\arcdeg$ & Position angle of source at 197\,MHz \tabularnewline
246 & residual\_mean\_197 & Jy\,beam$^{-1}$ & Mean of residual after source fitting at 197\,MHz \tabularnewline
247 & residual\_std\_197 & Jy\,beam$^{-1}$ & Standard deviation of residual after source fitting at 197\,MHz \tabularnewline
248 & psf\_a\_197 & $\arcsec$ & Major axis of PSF at location of source at 197\,MHz \tabularnewline
249 & psf\_b\_197 & $\arcsec$ & Minor axis of PSF at location of source at 197\,MHz \tabularnewline
250 & psf\_pa\_197 & $\arcdeg$ & Position angle of PSF at location of source at 197\,MHz \tabularnewline
251 & background\_204 & Jy\,beam$^{-1}$ & Background at 204\,MHz \tabularnewline
252 & local\_rms\_204 & Jy\,beam$^{-1}$ & Local RMS at 204\,MHz \tabularnewline
253 & peak\_flux\_204 & Jy\,beam$^{-1}$ & Peak flux density at 204\,MHz \tabularnewline
254 & err\_peak\_flux\_204 & Jy\,beam$^{-1}$ & Fitting error on peak flux density at 204\,MHz \tabularnewline
255 & int\_flux\_204 & Jy & Integrated flux density at 204\,MHz \tabularnewline
256 & err\_int\_flux\_204 & Jy & Fitting error on integrated flux density at 204\,MHz \tabularnewline
257 & a\_204 & $\arcsec$ & Major axis of source at 204\,MHz \tabularnewline
258 & b\_204 & $\arcsec$ & Minor axis of source at 204\,MHz \tabularnewline
259 & pa\_204 & $\arcdeg$ & Position angle of source at 204\,MHz \tabularnewline
260 & residual\_mean\_204 & Jy\,beam$^{-1}$ & Mean of residual after source fitting at 204\,MHz \tabularnewline
261 & residual\_std\_204 & Jy\,beam$^{-1}$ & Standard deviation of residual after source fitting at 204\,MHz \tabularnewline
262 & psf\_a\_204 & $\arcsec$ & Major axis of PSF at location of source at 204\,MHz \tabularnewline
263 & psf\_b\_204 & $\arcsec$ & Minor axis of PSF at location of source at 204\,MHz \tabularnewline
264 & psf\_pa\_204 & $\arcdeg$ & Position angle of PSF at location of source at 204\,MHz \tabularnewline
265 & background\_212 & Jy\,beam$^{-1}$ & Background at 212\,MHz \tabularnewline
266 & local\_rms\_212 & Jy\,beam$^{-1}$ & Local RMS at 212\,MHz \tabularnewline
267 & peak\_flux\_212 & Jy\,beam$^{-1}$ & Peak flux density at 212\,MHz \tabularnewline
268 & err\_peak\_flux\_212 & Jy\,beam$^{-1}$ & Fitting error on peak flux density at 212\,MHz \tabularnewline
269 & int\_flux\_212 & Jy & Integrated flux density at 212\,MHz \tabularnewline
270 & err\_int\_flux\_212 & Jy & Fitting error on integrated flux density at 212\,MHz \tabularnewline
271 & a\_212 & $\arcsec$ & Major axis of source at 212\,MHz \tabularnewline
272 & b\_212 & $\arcsec$ & Minor axis of source at 212\,MHz \tabularnewline
273 & pa\_212 & $\arcdeg$ & Position angle of source at 212\,MHz \tabularnewline
274 & residual\_mean\_212 & Jy\,beam$^{-1}$ & Mean of residual after source fitting at 212\,MHz \tabularnewline
275 & residual\_std\_212 & Jy\,beam$^{-1}$ & Standard deviation of residual after source fitting at 212\,MHz \tabularnewline
276 & psf\_a\_212 & $\arcsec$ & Major axis of PSF at location of source at 212\,MHz \tabularnewline
277 & psf\_b\_212 & $\arcsec$ & Minor axis of PSF at location of source at 212\,MHz \tabularnewline
278 & psf\_pa\_212 & $\arcdeg$ & Position angle of PSF at location of source at 212\,MHz \tabularnewline
279 & background\_220 & Jy\,beam$^{-1}$ & Background at 220\,MHz \tabularnewline
280 & local\_rms\_220 & Jy\,beam$^{-1}$ & Local RMS at 220\,MHz \tabularnewline
281 & peak\_flux\_220 & Jy\,beam$^{-1}$ & Peak flux density at 220\,MHz \tabularnewline
282 & err\_peak\_flux\_220 & Jy\,beam$^{-1}$ & Fitting error on peak flux density at 220\,MHz \tabularnewline
283 & int\_flux\_220 & Jy & Integrated flux density at 220\,MHz \tabularnewline
284 & err\_int\_flux\_220 & Jy & Fitting error on integrated flux density at 220\,MHz \tabularnewline
285 & a\_220 & $\arcsec$ & Major axis of source at 220\,MHz \tabularnewline
286 & b\_220 & $\arcsec$ & Minor axis of source at 220\,MHz \tabularnewline
287 & pa\_220 & $\arcdeg$ & Position angle of source at 220\,MHz \tabularnewline
288 & residual\_mean\_220 & Jy\,beam$^{-1}$ & Mean of residual after source fitting at 220\,MHz \tabularnewline
289 & residual\_std\_220 & Jy\,beam$^{-1}$ & Standard deviation of residual after source fitting at 220\,MHz \tabularnewline
290 & psf\_a\_220 & $\arcsec$ & Major axis of PSF at location of source at 220\,MHz \tabularnewline
291 & psf\_b\_220 & $\arcsec$ & Minor axis of PSF at location of source at 220\,MHz \tabularnewline
292 & psf\_pa\_220 & $\arcdeg$ & Position angle of PSF at location of source at 220\,MHz \tabularnewline
293 & background\_227 & Jy\,beam$^{-1}$ & Background at 227\,MHz \tabularnewline
294 & local\_rms\_227 & Jy\,beam$^{-1}$ & Local RMS at 227\,MHz \tabularnewline
295 & peak\_flux\_227 & Jy\,beam$^{-1}$ & Peak flux density at 227\,MHz \tabularnewline
296 & err\_peak\_flux\_227 & Jy\,beam$^{-1}$ & Fitting error on peak flux density at 227\,MHz \tabularnewline
297 & int\_flux\_227 & Jy & Integrated flux density at 227\,MHz \tabularnewline
298 & err\_int\_flux\_227 & Jy & Fitting error on integrated flux density at 227\,MHz \tabularnewline
299 & a\_227 & $\arcsec$ & Major axis of source at 227\,MHz \tabularnewline
300 & b\_227 & $\arcsec$ & Minor axis of source at 227\,MHz \tabularnewline
301 & pa\_227 & $\arcdeg$ & Position angle of source at 227\,MHz \tabularnewline
302 & residual\_mean\_227 & Jy\,beam$^{-1}$ & Mean of residual after source fitting at 227\,MHz \tabularnewline
303 & residual\_std\_227 & Jy\,beam$^{-1}$ & Standard deviation of residual after source fitting at 227\,MHz \tabularnewline
304 & psf\_a\_227 & $\arcsec$ & Major axis of PSF at location of source at 227\,MHz \tabularnewline
305 & psf\_b\_227 & $\arcsec$ & Minor axis of PSF at location of source at 227\,MHz \tabularnewline
306 & psf\_pa\_227 & $\arcdeg$ & Position angle of PSF at location of source at 227\,MHz \tabularnewline
307 & int\_flux\_fit\_200 & Jy & Fitted flux density at 200\,MHz \tabularnewline
308 & err\_int\_flux\_fit\_200 & Jy & Error on fitted flux density at 200\,MHz \tabularnewline
309 & alpha & -- & Fitted spectral index assuming a power-law SED \tabularnewline
310 & err\_alpha & -- & Error on fitted spectral index \tabularnewline
311 & reduced\_chi2 & -- & Reduced $\chi^2$ statistic for power-law SED fit \tabularnewline
\hline
\end{longtable}


\label{lastpage}

\end{document}